\documentclass[a4paper,11pt]{article}
\pdfoutput=1 

\usepackage{jinstpub} 
\usepackage[english]{babel}
\usepackage[TS1,T1]{fontenc}
\usepackage[utf8]{inputenc}
\usepackage[load=addn]{siunitx}
\usepackage[export]{adjustbox}
\usepackage{subfig}
\usepackage{lineno}
\usepackage{booktabs}
\usepackage[autolanguage]{numprint}

\renewcommand*\npstyleenglish{%
	\npthousandsep{\,}%
	\npdecimalsign{.}%
	\npproductsign{\cdot}%
	\npunitseparator{\,}%
	\npdegreeseparator{}%
	\npcelsiusseparator{\nprt@unitsep}%
	\nppercentseparator{\nprt@unitsep}%
}

\newcommand{\ohmcm}{\si{\ohm\cdot cm}}
\newcommand{\Neq}{\SI{1}{\mega\eV}~$\text{n}_{\text{eq}}/\text{{cm}}^2$}

\title{Charge collection characterisation with the Transient Current Technique of the ams H35DEMO CMOS detector after proton irradiation}

\author[a]{J.~Anders,}
\author[b]{M.~Benoit,}
\author[a]{S.~Braccini,}
\author[c]{R.~Casanova,}
\author[d]{H.~Chen,}
\author[d]{K.~Chen,}
\author[b]{F.A.~Di Bello,}
\author[a]{A.~Fehr,}
\author[b]{D.~Ferrere,}
\author[a]{D.~Forshaw,}
\author[b]{T.~Golling,}
\author[b]{S.~Gonzalez-Sevilla,}
\author[b]{G.~Iacobucci,}
\author[b]{M.~Kiehn,}
\author[d]{F.~Lanni,}
\author[d,e]{H.~Liu,}
\author[b,f]{L.~Meng,}
\author[a,1]{C.~Merlassino,\note{Corresponding author.}}
\author[a]{A.~Miucci,}
\author[b,g]{M.~Nessi,}
\author[h]{I.~Peri\'c,}
\author[a]{M.~Rimoldi,}
\author[b]{D~M~S~Sultan,}
\author[b]{M.~Vicente Barreto Pinto,}
\author[f]{E.~Vilella,}
\author[a]{M.~Weber,}
\author[a]{T.~Weston,}
\author[d]{W.~Wu,}
\author[d]{L.~Xu,}
\author[b,1]{and E.~Zaffaroni}

\affiliation[a]{Albert Einstein Center for Fundamental Physics and Laboratory for High Energy Physics, University of Bern, Siedlerstrasse 5, CH-3012 Bern, Switzerland}
\affiliation[b]{Département de Physique Nucléaire et Corpusculaire
	(DPNC), Université de Genève, 24 quai Ernest Ansermet 1211 Genève 4, Switzerland}
\affiliation[c]{Institut de Física d’Altes Energies (IFAE), The Barcelona Institute of Science and Technology, Edifici CN, UAB campus, 08193 Bellaterra (Barcelona), Spain}
\affiliation[d]{Brookhaven National Laboratory (BNL), P.O. Box 5000, Upton, NY 11973-5000, USA}
\affiliation[e]{Dept. of Modern Physics, University of Science and Technology of China, Hefei, Anhui 230026, China}
\affiliation[f]{Department of Physics, University of Liverpool, The Oliver Lodge Laboratory, Liverpool L69 7ZE, UK}
\affiliation[g]{European Organization for Nuclear Research (CERN), 385 route de Meyrin, 1217 Meyrin, Switzerland}
\affiliation[h]{Karlsruhe Institute of Technology (KIT), IPE, 76021 Karlsruhe, Germany}

\emailAdd{claudia.merlassino@cern.ch}
\emailAdd{ettore.zaffaroni@unige.ch}

\abstract{This paper reports on the characterisation with Transient Current Technique measurements of the charge collection and depletion depth of a radiation-hard high-voltage CMOS pixel sensor produced at ams AG. Several substrate resistivities were tested before and after proton irradiation with two different sources: the 24~GeV Proton Synchrotron at CERN and the 16.7~MeV Cyclotron at Bern Inselspital.}


\begin{document}
\maketitle
\flushbottom
	\section{Introduction}
\label{sec:intro}
In view of the High-Luminosity Large Hadron Collider (HL-LHC)~\cite{hlLHC}, the ATLAS experiment is planning an upgrade of its tracker~\cite{Collaboration:2257755}, which is expected to operate for more than ten years at high fluence, to collect a dataset of 3000~\(\text{fb}^{-1}\). The new Inner Tracker (ITk) will consist entirely of silicon detectors, with the barrel section being divided in five layers of pixel detectors, close to the beam line, with four additional strip layers surrounding it.
In this context, it is interesting to explore new kinds of technologies for silicon sensors, which are able to withstand the high radiation environment of the upgraded accelerator and allow to take advantage of the large-scale industrial expertise. 

One of the pixel detector options considered for the ITk is the use of a High-Voltage CMOS (HV-CMOS) technology, an industry standard, to build a depleted monolithic sensor. This technology offers the possibility to reduce the pixel size, the power consumption and the material budget. Additionally, HV-CMOS sensors will have a simpler assembly procedure and a potential better yield compared to hybrid detectors.
This technology is available in different resistivities and allows for a depletion depth from few tens to more than a hundred microns~\cite{PERIC2007876}. In these sensors charge is mostly collected by drift, so these devices provide faster and larger signals with respect to the common undepleted monolithic sensors, where charge is collected mainly by diffusion, and feature a higher level of radiation tolerance.

In this paper the charge collection and the depletion depth of a HV-CMOS prototype~\cite{h35demo}, produced at ams AG~\cite{ams} in its H35 technology, with 350~nm feature size, is studied with the Transient Current Technique (TCT) before and after proton irradiation. The sensor and the experimental setup are outlined in Section~\ref{sec:h35demo} and~\ref{sec:setup} respectively, the measurements, analysis and irradiation methods are described in Sections~\ref{sec:non-irr},~\ref{sec:irr} and~\ref{sec:char-irr}. Conclusions are presented in Section~\ref{sec:conclusions}.

\section{The H35DEMO ASIC}
\label{sec:h35demo}
\begin{figure}[h]
	\centering
	\includegraphics[width=\textwidth]{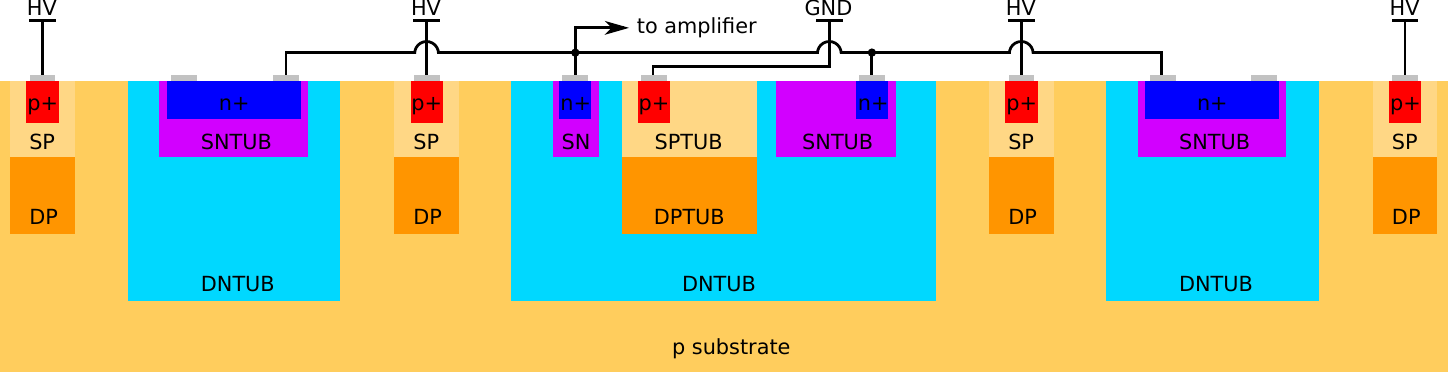}
	\caption{Simplified cross section of a test structure pixel. It is split into three n-wells connected together and the sensor pn junction is formed between the deep n-well (DNTUB) and the p substrate.}
	\label{fig:pixel-xsection}	
\end{figure}

The H35DEMO ASIC is a chip produced in the 350~nm technology at ams~AG. The area of the chip is $18.49\times 24.40$ \si{mm^2} and has been produced on substrates with four different nominal resistivities: 20, 80, 200 and 1000~\ohmcm{}.
The pixels of this ASIC have an area of $50\times 250$ \si{\micro m^2} and, in order to reduce the sensor capacitance while keeping a large fill factor, the electrode is split into three n-wells (shown in Figure~\ref{fig:pixel-xsection}). The pn junction of the sensor is between the p-doped substrate and the deep n-well.
The high voltage is provided by a dedicated pad and it is delivered from the top of the chip with p+ implants passing in between each n-well.

\begin{figure}[h]
	\centering
	\includegraphics[width=.7\textwidth]{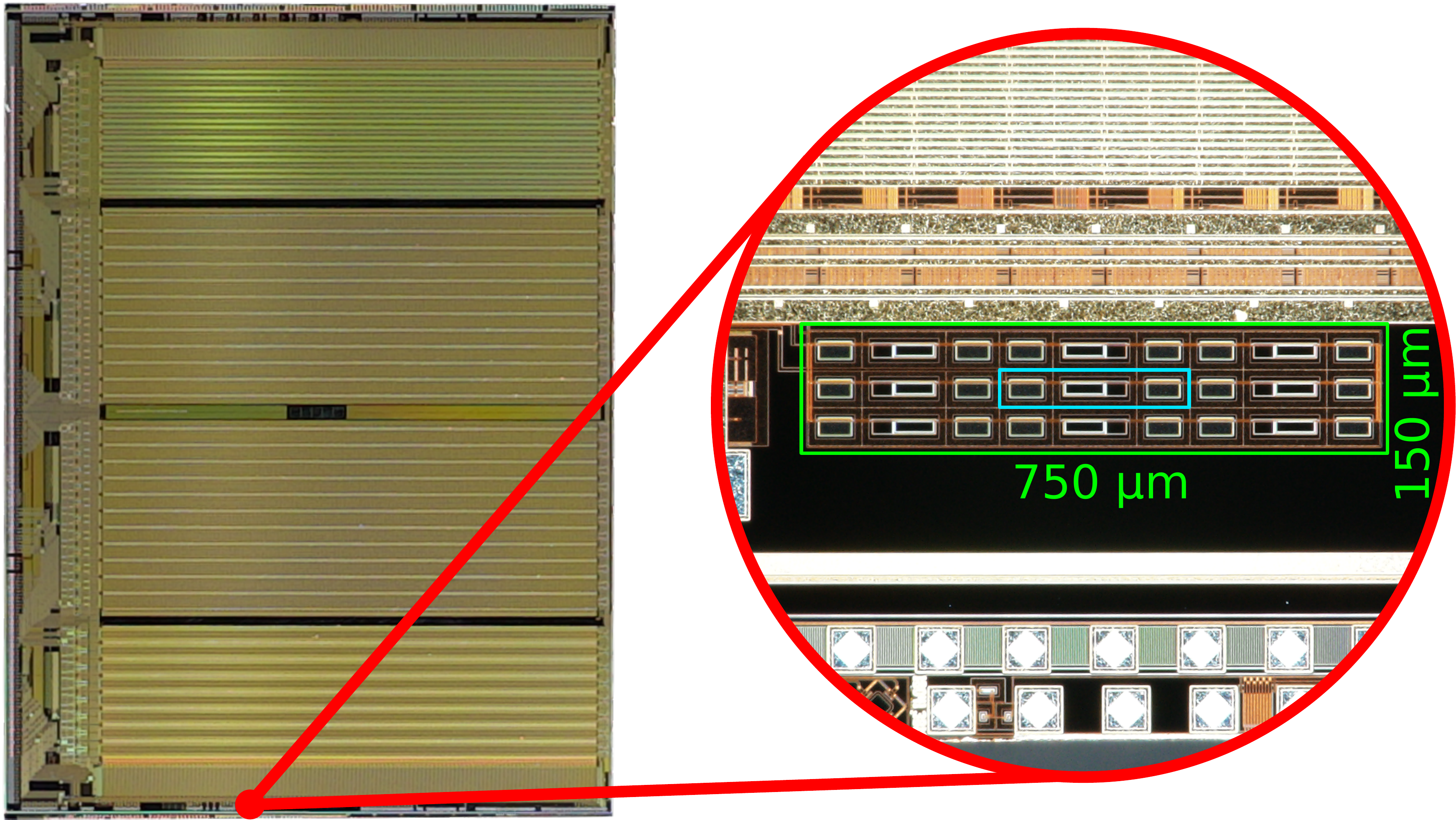}
	\caption{The H35DEMO and the test structure used for TCT measurements (encircled in red). The test structure consists of $3\times 3$ pixels (inside the large green rectangle) carrying no electronics. The central pixel (inside the small blue rectangle) is read out separately from the others.}
	\label{fig:h35demo}	
\end{figure}

In order to characterise the bulk properties, the chip contains several test structures on one edge. The structure used in this work is a $3\times 3$ pixel matrix (circled in Figure~\ref{fig:h35demo}), with no electronics, where the n-wells of the pixel are connected directly to a pad: the central pixel (circled in blue) and the surrounding ones are connected to two different lines, so it is possible to measure the charge collected by each of them separately.

%
%

\section{Experimental setup}
\label{sec:setup}
The depletion depth of the tested sensors has been measured with a Transient Current Technique, by illuminating the edge of the sensor with an infrared laser beam to generate electron-hole pairs inside the silicon bulk. The laser position with respect to the sensor is controlled using a micrometric $xyz$ stage system. The current pulse induced by the laser is amplified, digitised and integrated to estimate the total charge collected inside the sensor. Two-dimensional scans in the plane perpendicular to the laser beam are performed to measure the depletion depth.

\begin{figure}[h]
	\centering\includegraphics[width=\textwidth]{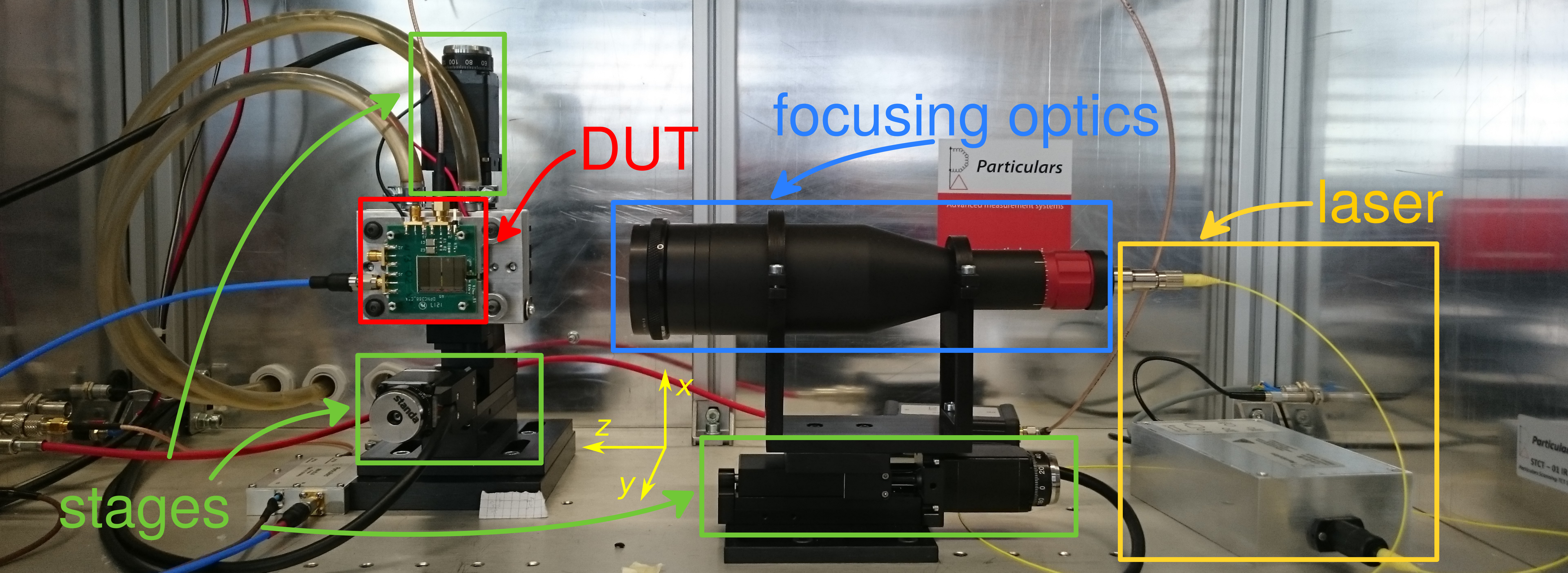}
	\caption{A picture of the TCT setup. The laser beam is supplied to the optics through an optical fiber and focalised on the device under test (DUT). Three stages move the DUT (along $x$ and $y$) and the focusing optics (along $z$) with 1~\si{\micro m} precision.}
	\label{fig:tct-setup}	
\end{figure}

The TCT measurements were carried out at the University of Bern and at the University of Geneva with two setups produced by Particulars~\cite{particulars} (see Figure~\ref{fig:tct-setup}), each consisting of three moving stages, two used to position the Device Under Test (DUT) and a third to adjust the focus of the laser. They can be positioned with 1~\si{\micro m} precision. In our coordinate system $y$ is the direction perpendicular to the DUT surface, $z$ is the direction of the laser beam and $x$ is perpendicular to $y$ and $z$.
A laser with a wavelength of 1064~nm, generating pulses $\sim 500$~ps long with a rate of 1~kHz, is focused using dedicated optics. The laser also provides the trigger signal for the oscilloscope. The output signals of the detector are read with two 53~dB current amplifiers~\cite{particulars}, digitised with a DRS4 oscilloscope~\cite{psi-drs} and acquired with a dedicated data acquisition software.

The DUT is cooled with a Peltier cell and a chiller: the temperature is controlled by software and dry air is supplied to the setup to avoid the formation of dew or frost.

\begin{figure}[t]
	\centering
	\subfloat[][\emph{}]
	{\includegraphics[width=.45\textwidth,valign=c]{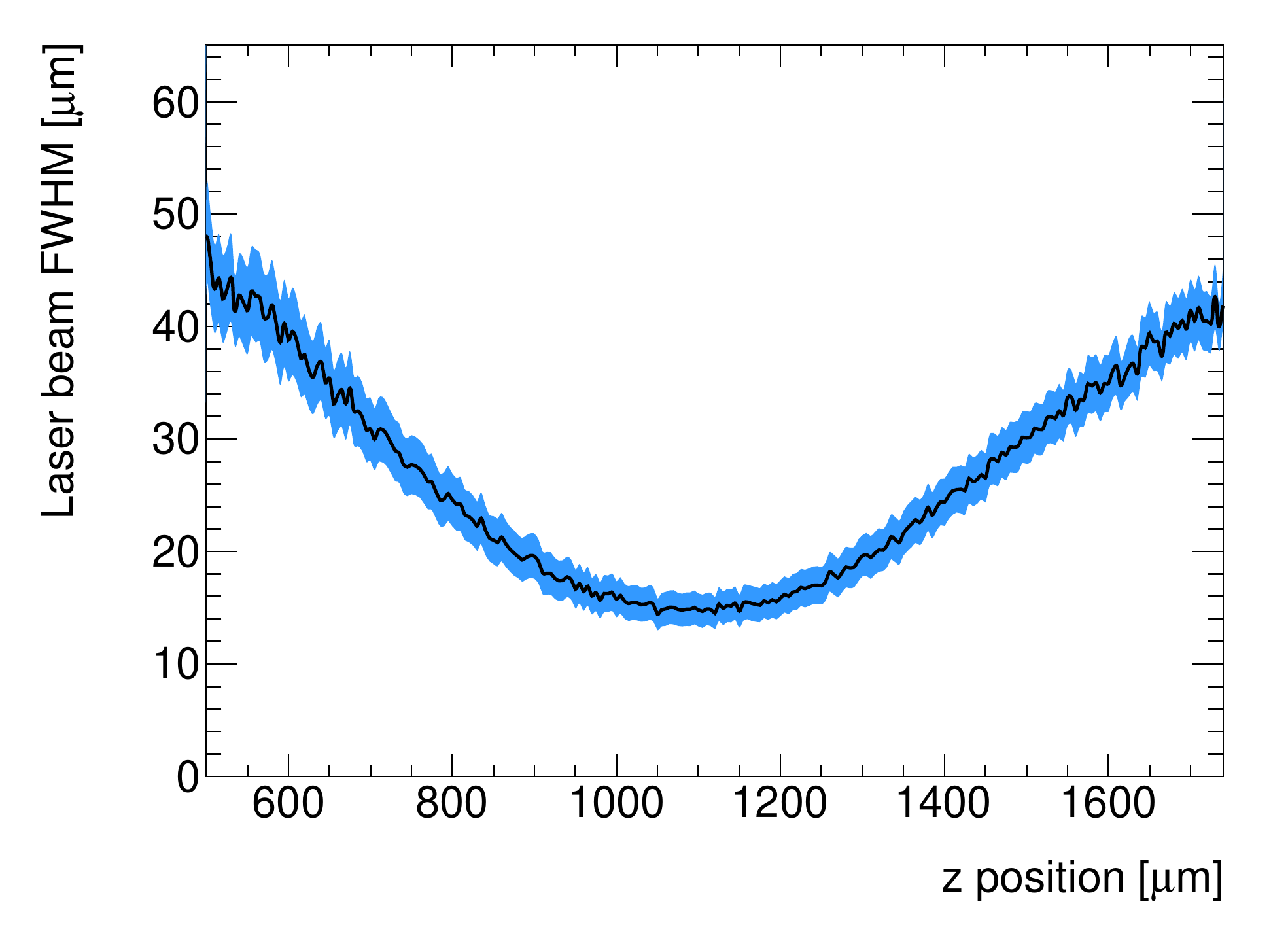} \label{plot:laser-width}} \quad
	\subfloat[][\emph{}]
	{\includegraphics[width=.45\textwidth,valign=c]{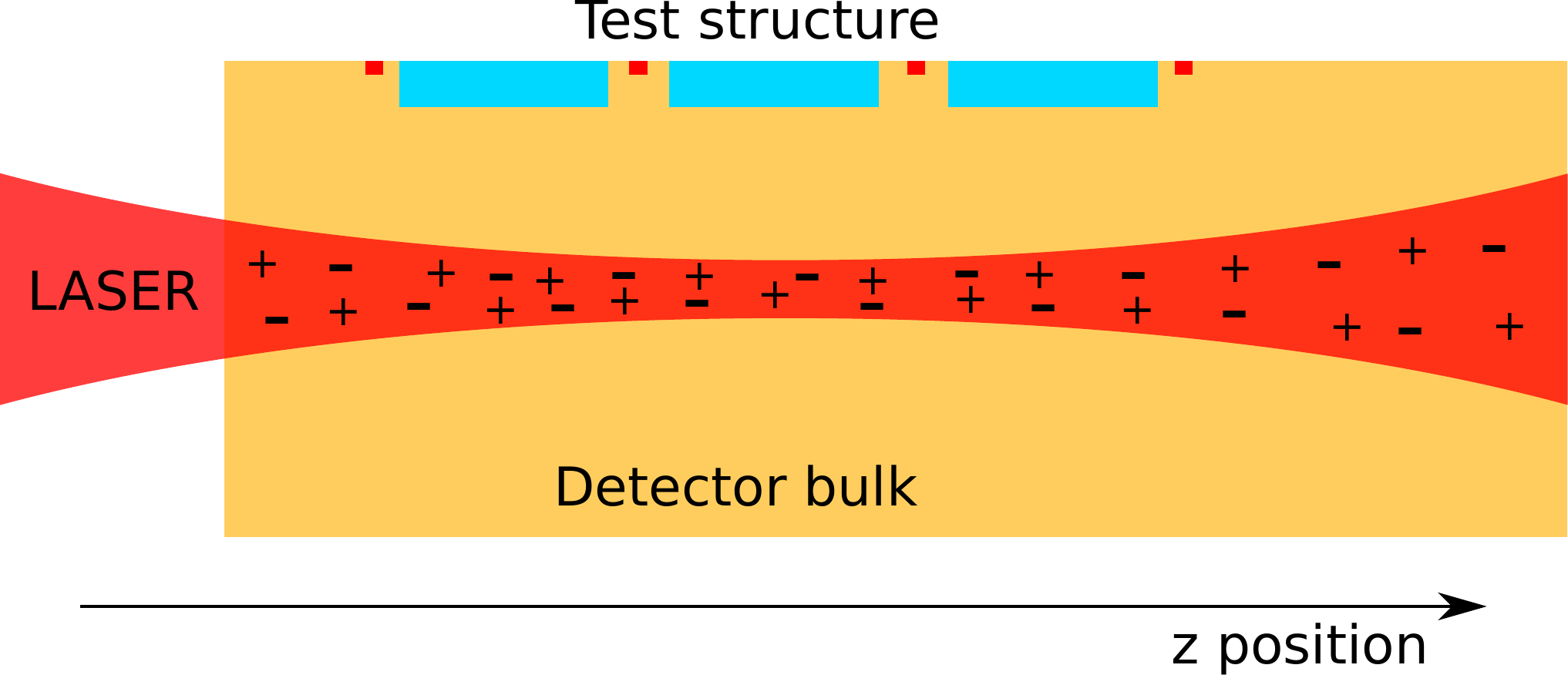} \label{fig:tct-laser}}
	\caption{(a) The FWHM of the laser beam around the focus. The region around the minimum is the best focus achievable and the aim is to position it exactly below the test structures. (b) A sketch of the charge generation inside the silicon: charge is generated along the laser beam, not just in the focus point.}

\end{figure}

The laser has a gaussian profile, which was measured by performing a scan perpendicular to the silicon-metal edge of a test sensor. The charge profile of this scan is modelled with an error function, whose $\sigma$ parameter is used to estimate the laser beam width. This procedure is repeated at different focus positions ($z$ direction) and the results are shown in Figure~\ref{plot:laser-width}.
The laser profile has a minimum Full Width at Half Maximum (FWHM) of $15\pm2$~\si{\micro m}.

The charge is generated along this profile (Figure~\ref{fig:tct-laser}), for this reason it is important to focus the laser before taking measurements. The laser width also influences the minimum depletion depth that can be measured.

A dedicated PCB has been developed for the measurements to guarantee a low-noise and clean signal.
The PCB features signal lines with controlled impedance and 50~\si{\ohm} resistors termination, a RC HV filter and a PT-100 thermistor placed a few millimetres from the sensor to record its temperature during the measurement. 
	\section{Characterisation of the non irradiated samples}
\label{sec:non-irr}
\begin{figure}[h]
	\centering
	\subfloat[][\emph{}]
	{\includegraphics[width=.45\textwidth]{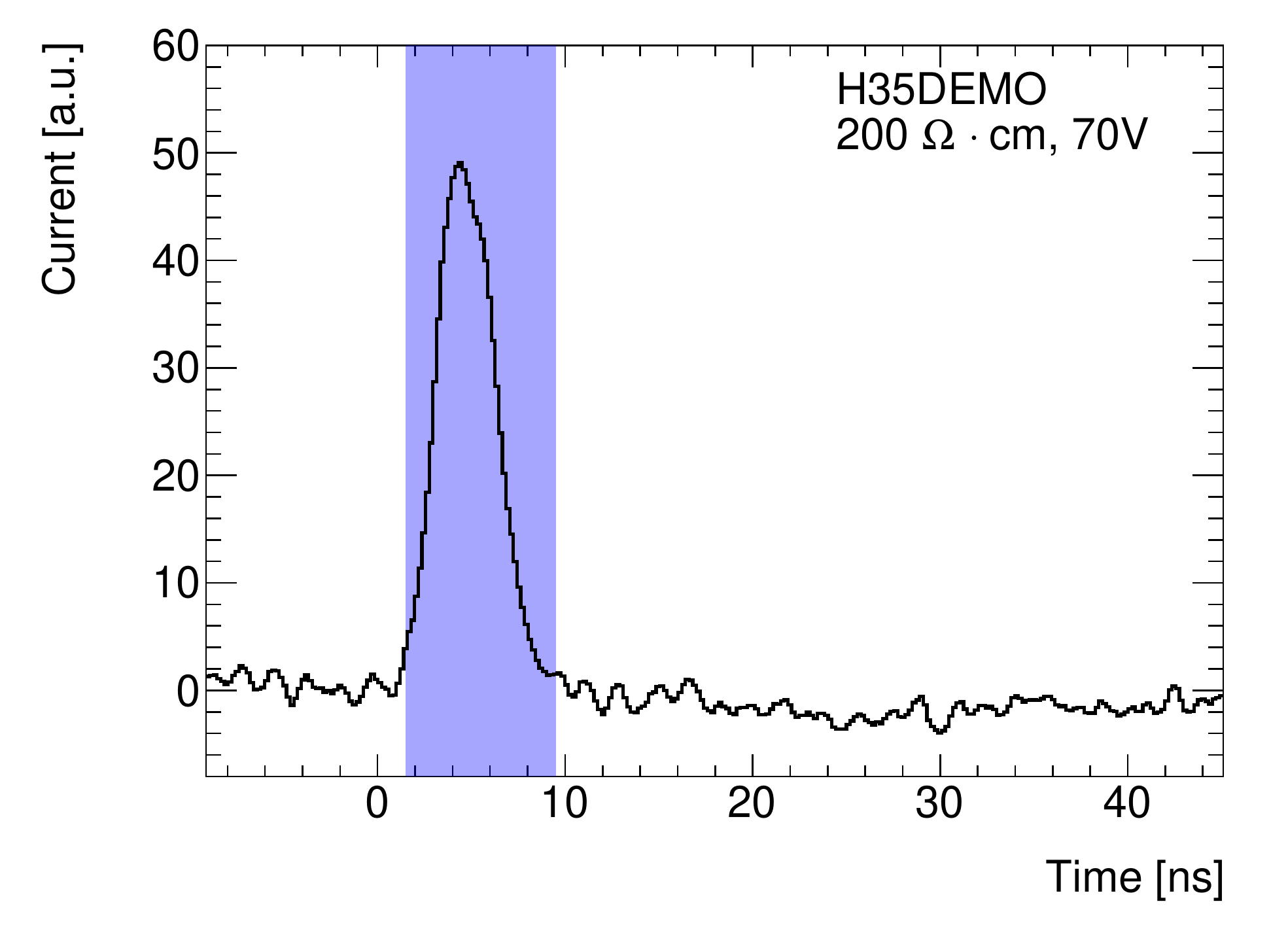} \label{plot:wf}} \quad
	\subfloat[][\emph{}]
	{\includegraphics[width=.45\textwidth]{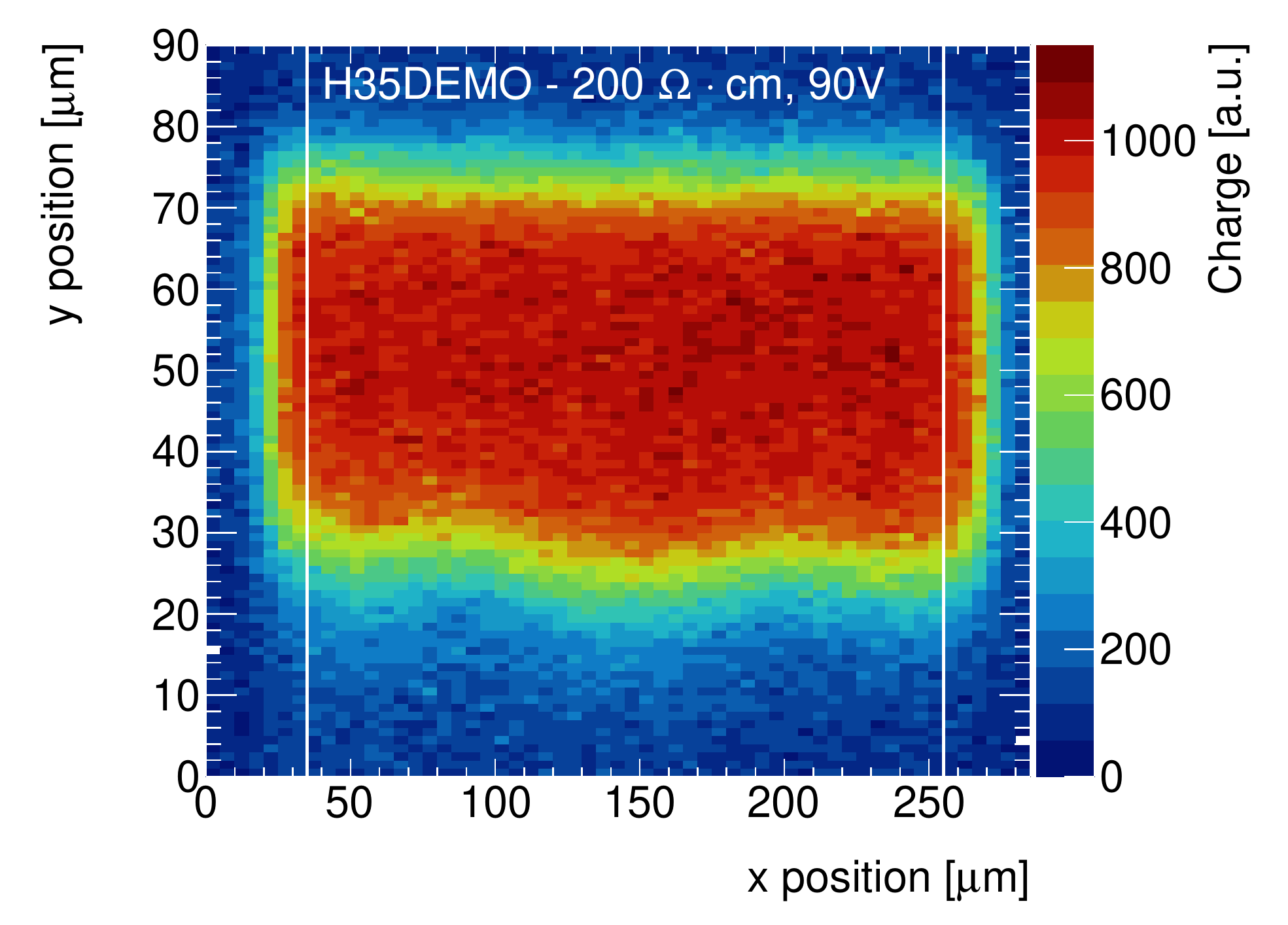} \label{plot:2d}} 
	\caption{(a) An example of an amplified current waveform collected in the depletion region. The vertical band highlights the integration window used to measure the collected charge. (b) Collected charge as a function of the position of the laser focus for a 200~\ohmcm{} sample at $90$~V bias voltage. The vertical lines represent the range used to estimate the depletion region.}
	
\end{figure}

Edge TCT measurements were carried out to estimate the depletion depth as a function of the bias voltage on all available substrate resistivities. After focusing the laser, a 2-dimensional scan was taken at different reverse bias voltages (0 to $100$~V, in 10~V steps). The scan was performed with 1~\si{\micro m} steps in the $y$ direction, from the implants down inside the bulk, and 5~\si{\micro m} steps in the $x$ direction, parallel to the long edge of the pixels. For each point of the scan, a full current waveform  (obtained after averaging 40 measurements to decrease noise) was saved. An example of the current waveform inside the depletion region is shown in Figure~\ref{plot:wf}. To obtain the collected charge, each waveform was integrated over an 8 ns time window, highlighted as a vertical band in the plot. An example of the result, showing a 2D histogram of the charge collection map, is presented in Figure~\ref{plot:2d}: each bin of the 2D histogram shows the integral of the waveform recorded while pointing the laser at the indicated \((x,y)\) coordinate.

\begin{figure}[t]
\centering
\subfloat[][\emph{}]
   {\includegraphics[width=.45\textwidth]{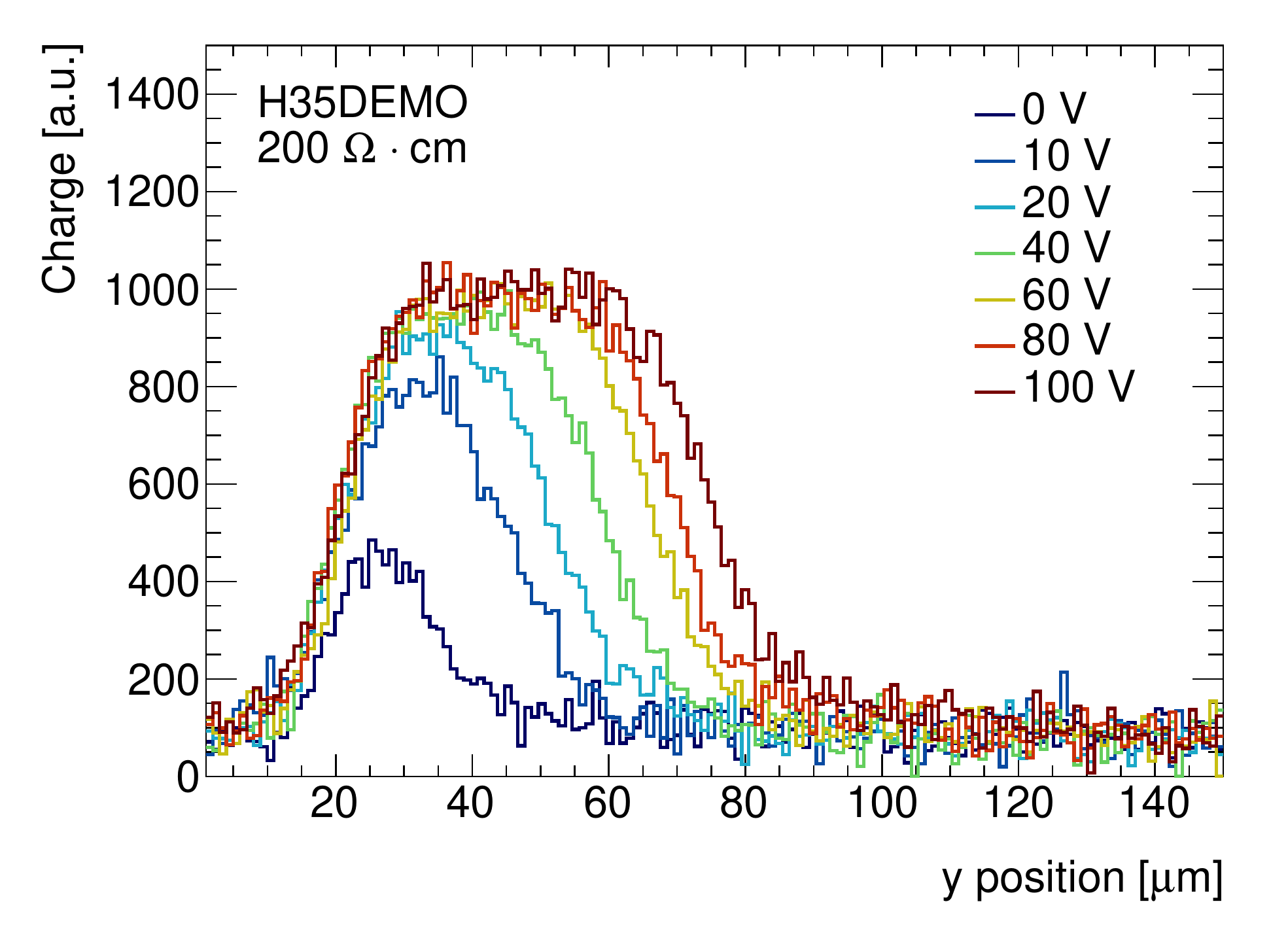} \label{plot:projection-200}} \quad
\subfloat[][\emph{}]
   {\includegraphics[width=.45\textwidth]{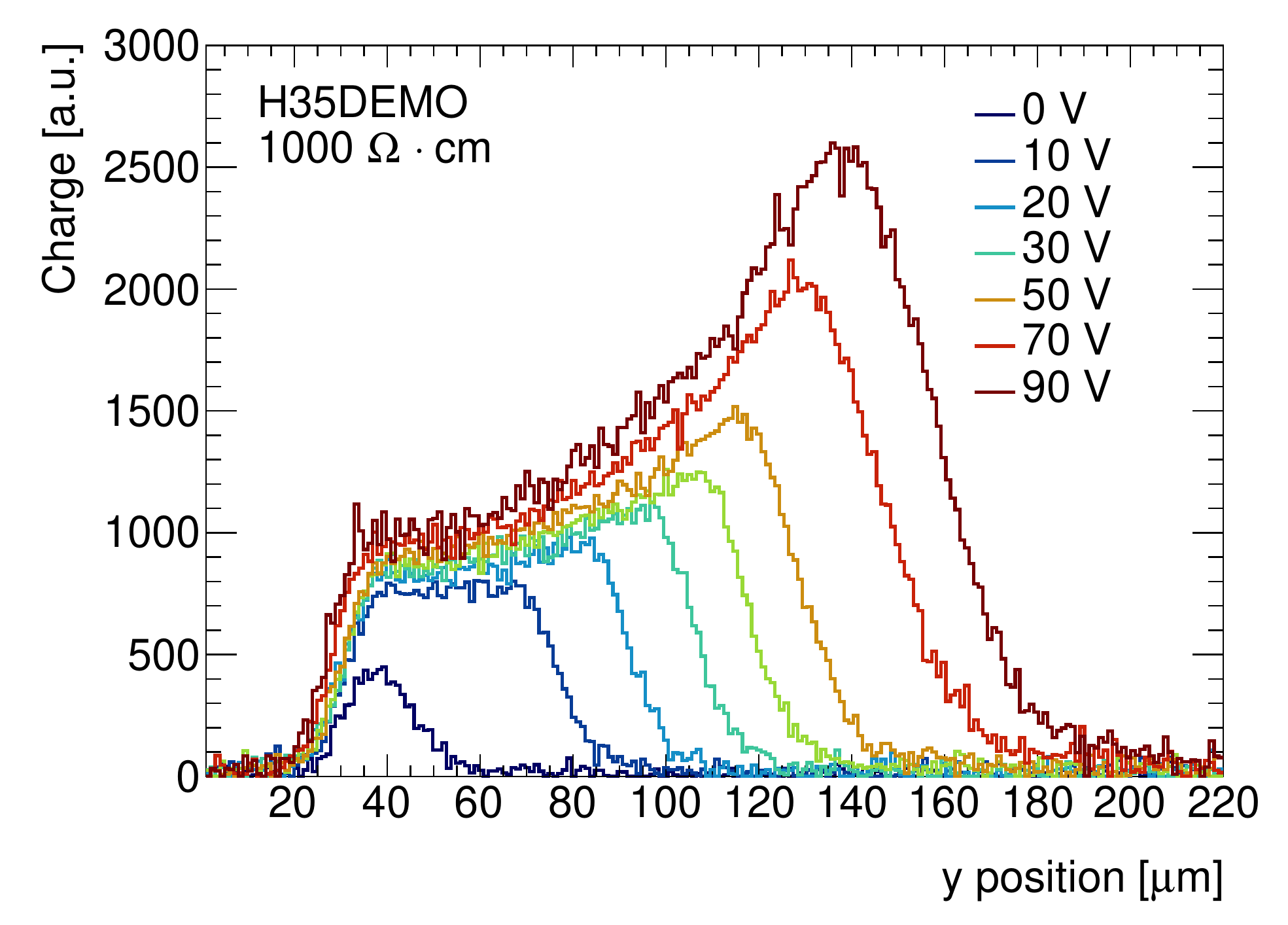} \label{plot:projection-1000}} 
   \caption{Projections, perpendicular to the sensor surface, of the collected charge for (a) a 200~\ohmcm{} sample and (b) for a 1000~\ohmcm{} sample.}
	\label{plot:projection}
\end{figure}

To estimate the depletion depth from these measurements it is necessary to take the profile along the $y$ direction of the collected charge distribution and then calculate the FWHM. Examples of these profiles, at different bias voltages, are shown in Figure~\ref{plot:projection}.

To have a good estimation, it is necessary to remove the profiles which are too close to the lateral edge of the pixel, where the laser beam is not fully contained. The average charge of each profile is calculated and if it is smaller that 0.85 times the maximum average charge, it is not considered. The profiles selected in this way are between the white vertical lines in Figure~\ref{plot:2d} and designate the fiducial range.

In case of the 1000~\ohmcm{} sample (Figure~\ref{plot:projection-1000}), a peak appears in the otherwise flat charge collection profile when the reverse bias voltage exceeds $30$~V. This is due to the shape of the electric field lines inside the sensor bulk: when electron-hole pairs are generated  deeper inside the bulk (depletion depth greater than $\sim80$~\si{\micro m}), part of the charge created below the outermost pixels of the test structure is collected by the central pixel, producing the peak in the charge collection profile.
This is confirmed by the charge collection profile of the pixel frame, which features a dip corresponding to the charge excess observed in the central pixel. A flat charge collection profile is obtained by summing the two.

The peak produced by the charge generated in the neighbouring pixels can produce a bias in the evaluation of the FWHM~\cite{ifae-tct}. Therefore, to separate the different contributions, each profile is fitted with the following function:

\begin{equation}
	f(y) = \frac{A}{4}\left[ \mathrm{erf}\left(\frac{y-y_1}{\sigma_1}\right) +1 \right] \cdot \mathrm{erfc} \left( \frac{y-y_2}{\sigma_2}\right) + B\cdot \mathrm{Gaussian}\left(y, \mu, \sigma\right) + C
	\label{eq:fit}
\end{equation}

\begin{figure}[h]
	\centering
	\subfloat[][\emph{}]
	{\includegraphics[width=.45\textwidth]{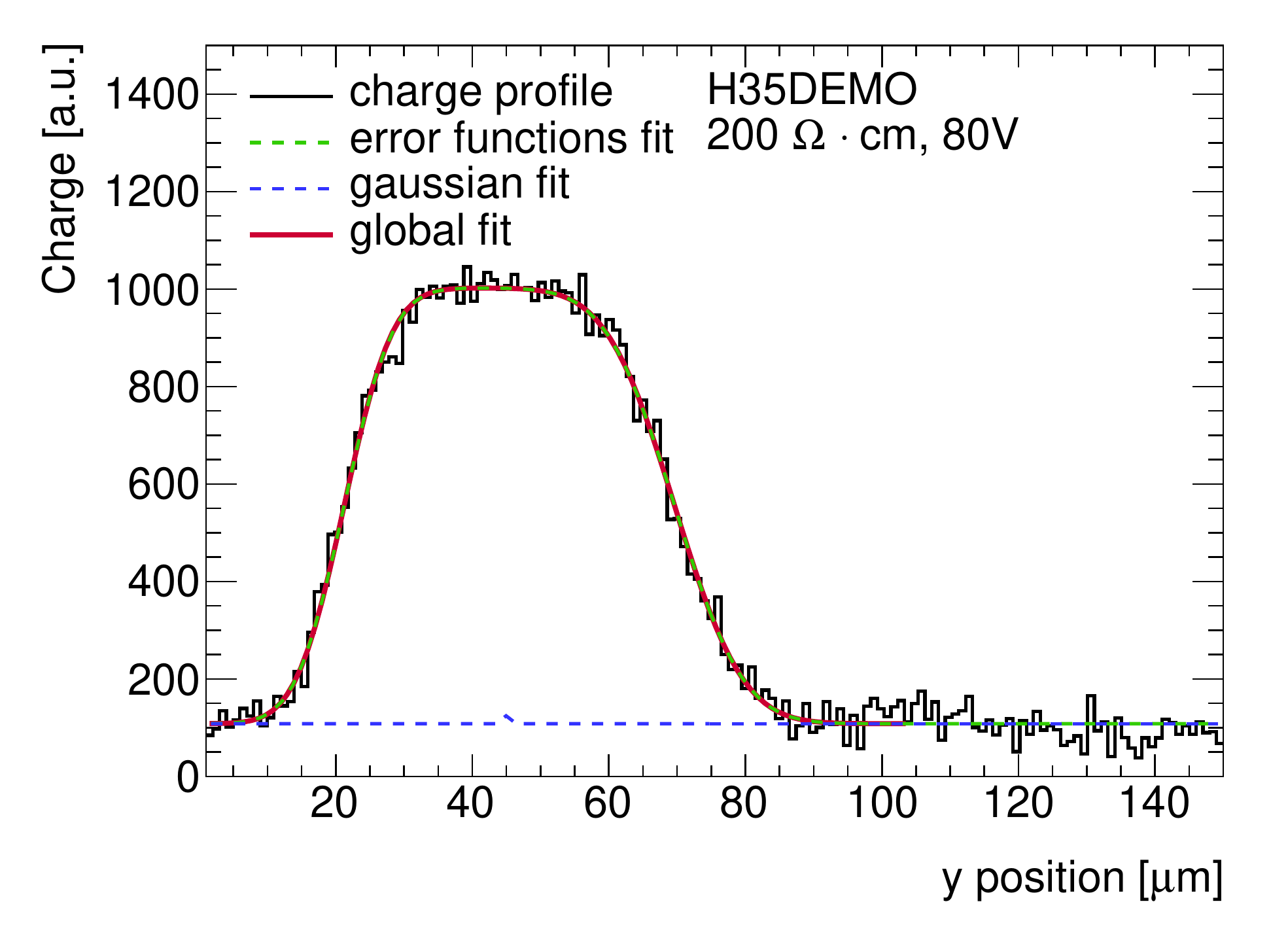} \label{plot:fit-200}} \quad
	\subfloat[][\emph{}]
	{\includegraphics[width=.45\textwidth]{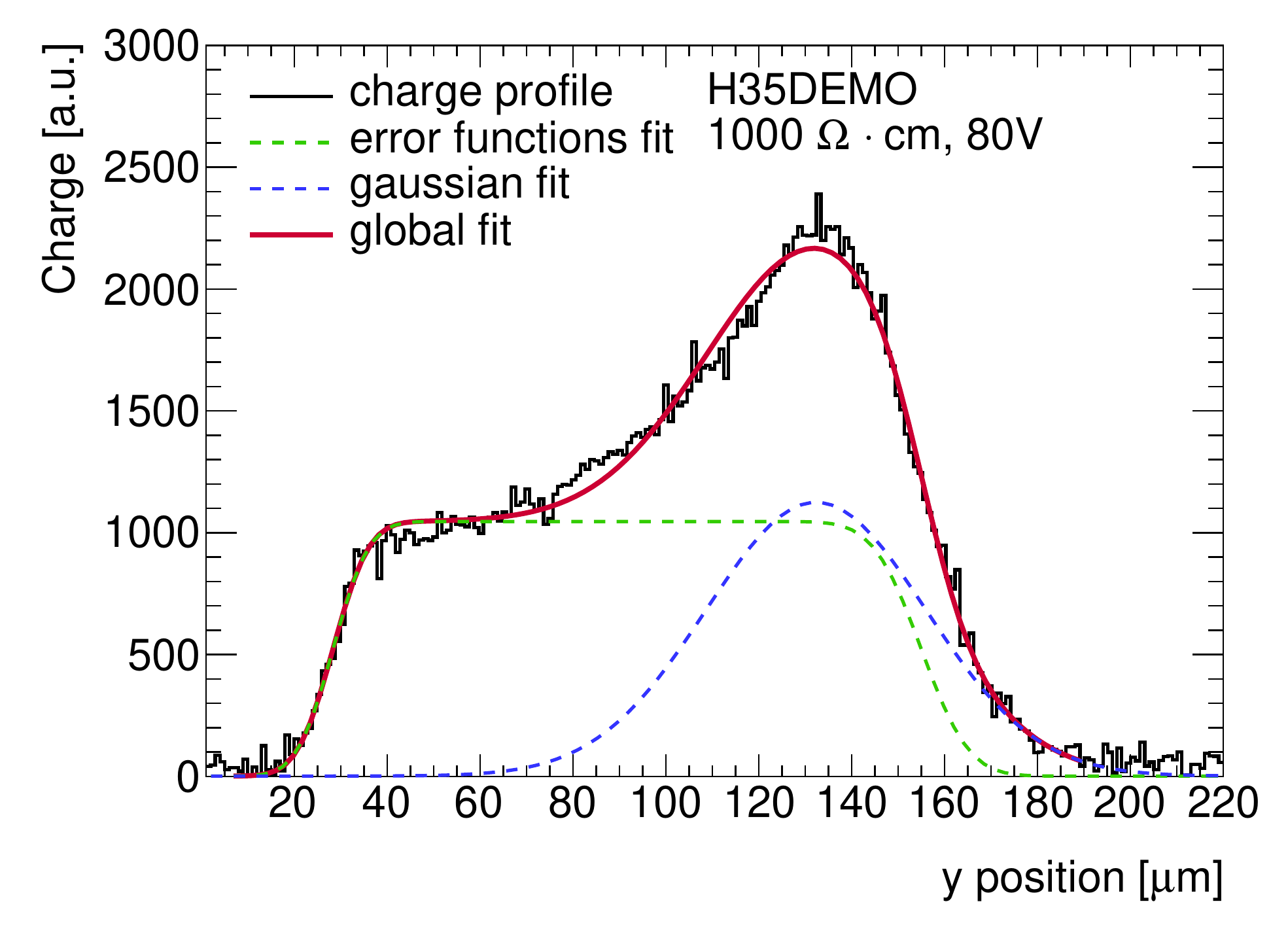} \label{plot:fit-1000}} 
	\caption{Examples of the charge profile fitted with the function defined in Equation~\ref{eq:fit}, for a 200~\ohmcm{} sample (a) and for a 1000~\ohmcm{} sample (b). In the former, the charge sharing is negligible and the only contributions to the fit come from the error functions, while in the latter the charge sharing contribution is modelled with a Gaussian.}
	\label{plot:fit}
\end{figure}

The combination of the error functions represents the profile of the charge collected in the depletion region: \(y_1\) and \(y_2\) are the beginning and the end of the region respectively, while \(\sigma_1\) and \(\sigma_2\) describe the smearing of the charge profile due to the size of the beam spot, and, in the case of $\sigma_2$, the effect of the decreasing electric field inside the bulk. 
The gaussian (and its normalisation $B$) represents the position, the width and the amplitude of the peak due to charge sharing and $C$ a possible offset of the baseline current. The parameter $A$, which normalises the error functions and represents the charge collection maximum in absence of charge sharing, is used for the FWHM estimation.
When the depletion depth is comparable with the laser beam width, the charge collection profile has a gaussian shape and in this case the maximum is obtained from the global fitted function.
An example of the fit behaviour on two the different kind of profiles is presented in Figure~\ref{plot:fit}.

Once the maximum is calculated, the FWHM of the global fit profile is extracted. This procedure is repeated for each $x$ position in the fiducial range: the depletion depth is obtained from the average of the distribution of these values. The error on each point is evaluated combining the contribution of the charge sharing and the standard deviation of the formerly mentioned values.

This estimation of the depletion depth is not accurate when the depth is comparable to, or smaller than, the laser width at its focus: in those cases, since the charge collection profile is a convolution of the laser shape and of the depletion depth, when the latter is small, the estimation is dominated by the laser width.
To correct for this effect, the proportionality between the fraction of the laser contained in the depleted region and the total charge collected inside the sensor is exploited. Each depletion depth value is thus reweighted with the ratio obtained from the charge maximum obtained for each bias voltage (parameter $A$ from Equation~\ref{eq:fit}), divided by the charge maximum obtained when the laser is fully contained.

\begin{figure}[h]
	\centering
	{\includegraphics[width=.7\textwidth]{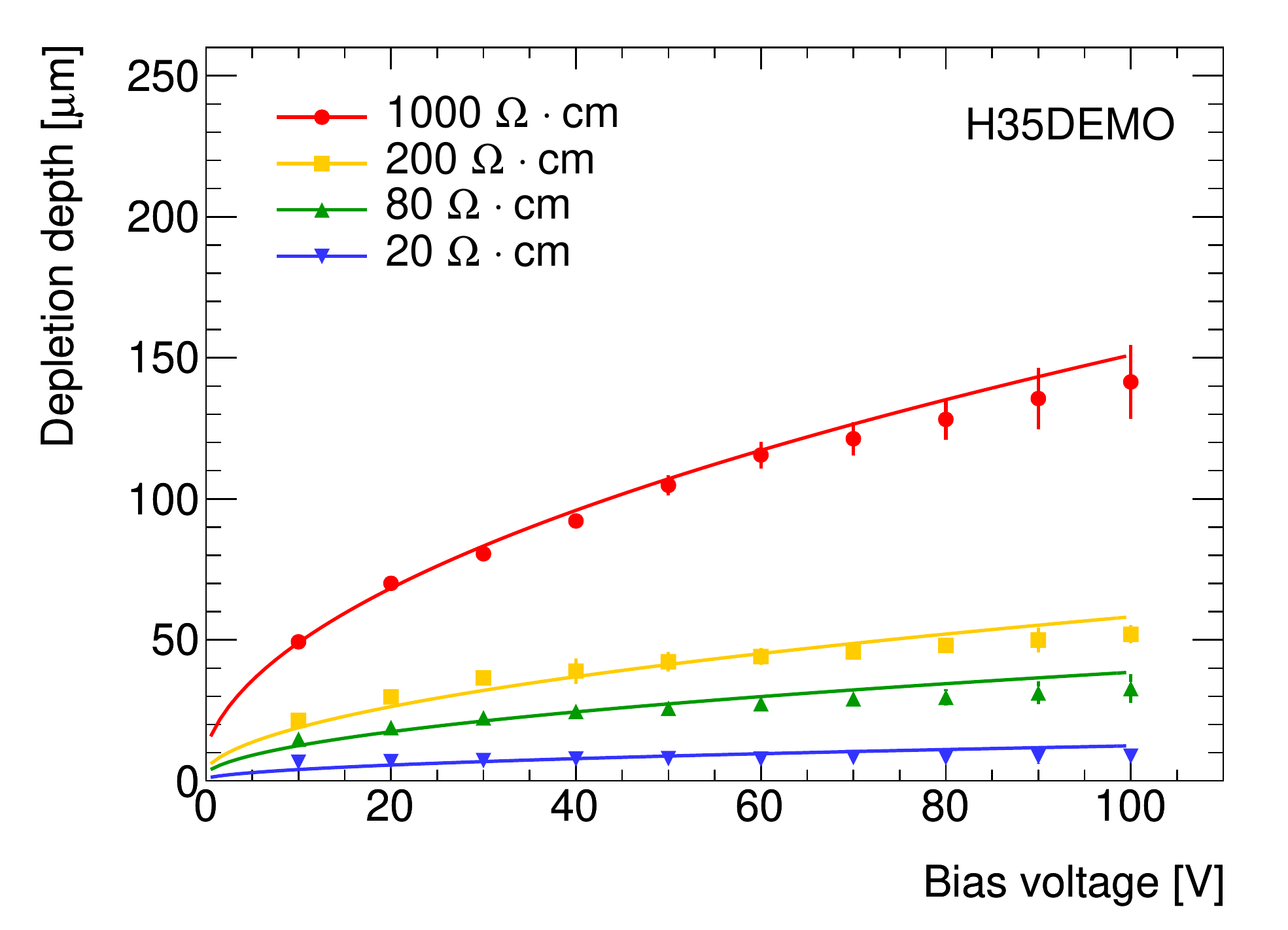}} 
	\caption{Evolution of the depletion depth as a function of the bias voltage. The four curves show the behaviour of samples with different resistivity: 20, 80, 200 and 1000 \ohmcm{}. The dots represent the measured points, while the lines are the fitted curves according to Equation~\ref{eq:rhofit}.}
	\label{plot:voltage_evolution}
\end{figure}

Figure~\ref{plot:voltage_evolution} shows the evolution of the depletion depth measured in this way, as a function of the bias voltage. The depletion depth as a function of the bias voltage is fitted with the following function in order to extract the resistivity of the substrate:

\begin{equation}
	d(V) = 0.3 \left[\si{\micro m}/\sqrt{\si{\ohm \cdot cm \cdot V}}\right] \cdot \sqrt{\rho(V+V_\text{bi})}
	\label{eq:rhofit}
\end{equation}
where $d$ is the depletion depth, $\rho$ is the substrate resistivity, $V$ the bias voltage and $V_\text{bi}$ the built-in voltage \cite{pdg-2014}. Since the value of $V_\text{bi}$ is measured for non irradiated samples, this parameter has been constrained in the range $0.2-0.6$~V in the fit. The fitted values are reported in Table~\ref{tab:fit-rho} and the fitted curves in Figure~\ref{plot:voltage_evolution}.

\begin{table}[h]
	\centering
	\caption{Resistivity values obtained from the fit, along with the resistivity ranges of the wafers declared by ams AG. The declared resistivity values are the ranges of resistivities provided by the manufacturer, while the nominal resistivity is the value used to identify a given range. }
	\label{tab:fit-rho}
	\begin{tabular}{ccc}
		\toprule
	Nominal $\rho$	& Declared $\rho$	& Fitted $\rho$ \\ 
	$[\ohmcm{}]$	& $[\ohmcm{}]$	& $[\ohmcm{}]$ 	\\ 
	\midrule
	20	& $15-25$	& $17\pm2$ \\ 
	\midrule
	80	& $50-100$	& $112\pm8$	\\ 
	\midrule
	200	& $200-400$	& $306\pm11$	\\
	\midrule
	1000& $600-1100$	& $2518\pm25$	\\
	\bottomrule
	\end{tabular}
\end{table}

The resistivity values obtained from the fit are consistent with the tolerance declared by ams for the 20, 80 and 200~\ohmcm{} samples, despite the fact that the formula used for the fit is an approximation and assumes the sensors to be back-biased. The estimated resistivity of the 1000~\ohmcm{} sample is not compatible with the declared resistivity values, but this behaviour has already been observed in other measurements~\cite{ifae-tct}.

\section{Irradiation campaign}
\label{sec:irr}
The sensors have been irradiated with \SI{16.7}{\mega\eV} and \SI{24}{\giga\eV} protons, in order to study the effect of the damage induced by charged hadrons. This kind of damage is particularly interesting because it is expected to be the major contribution to the total radiation damage in the outermost layers of the ATLAS tracker for the HL-LHC upgrade~\cite{Collaboration:2257755}.

The charged particles non-ionising energy loss (NIEL), mainly due to Coulomb interactions, is responsible for the displacement of atoms, producing defects in the lattice and altering the properties of charge collection inside the sensor. The amount of non-ionising energy released in the silicon depends on the type and the energy of the particles considered. For this reason, to compare the damage due to different particles or different energies, the NIEL is expressed as the fluence of 1~MeV neutrons. The proton fluences are thus reweighted with the appropriate hardness factor, taken to be 3.6 for the \SI{16.7}{\mega\eV} protons~\cite{HUHTINEN1993580,SR-NIEL} and 0.6 for the \SI{24}{\giga\eV} protons~\cite{psfactor}.

A first set of sensors was irradiated at the Bern cyclotron~\cite{Scampoli:2011zz}, which offers a low energy proton beam for medical applications and research purposes. The energy of the protons measured after the beam extraction window used during the irradiation is \SI{16.7}{\mega\eV}. The amount of dose delivered has been monitored during the irradiation by reading the current generated from the proton beam on collimator plates, located in vacuum, and an additional check is performed offline through the employment of dosimetric films~\cite{cyclotron}. The samples have been irradiated up to a fluence of \numprint{19e14}~\Neq, in several steps.

The second set of samples was irradiated at the CERN PS east area IRRAD facility~\cite{Ravotti:1951308}. PS IRRAD operates with \SI{24}{\giga\eV} protons, delivered from the PS accelerator in spills of about 400~ms (slow extraction), at a frequency of \(\sim\SI{0.1}{\Hz}\). The online monitoring of the dose is provided by a copper-based beam profile monitor~\cite{Gkotse:2235836}. An additional offline measurement of the dose is performed through \(\gamma\)-spectroscopy of thin aluminium films irradiated together with the samples~\cite{CURIONI2017101}. 
In this case, the samples were exposed to a fluence up to \numprint{20e14}~\Neq.

\section{Characterisation of the irradiated samples}
\label{sec:char-irr}

\begin{figure}[t]
\centering
\subfloat[][\emph{}]
   {\includegraphics[width=.45\textwidth]{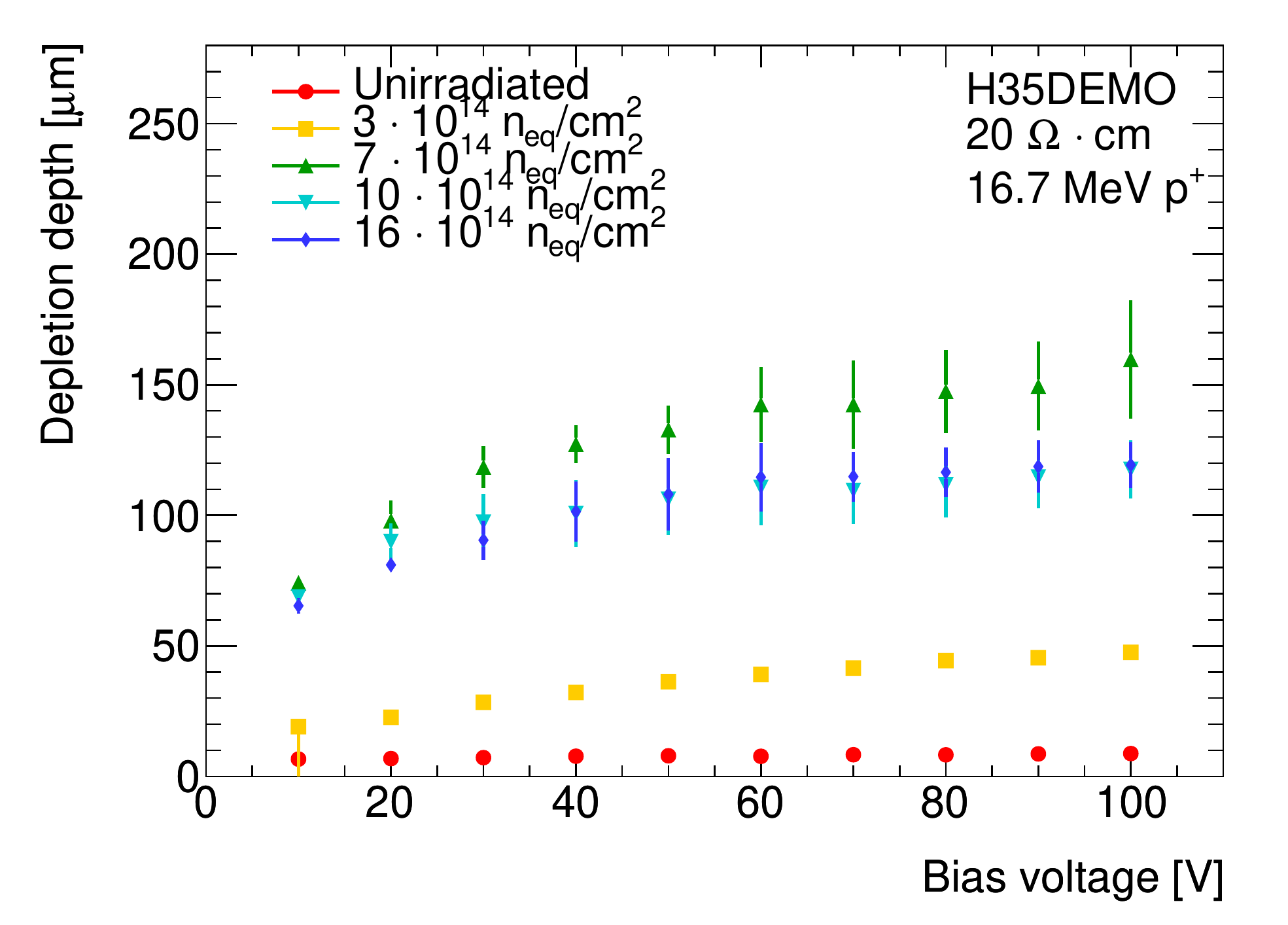}} \quad
\subfloat[][\emph{}]
   {\includegraphics[width=.45\textwidth]{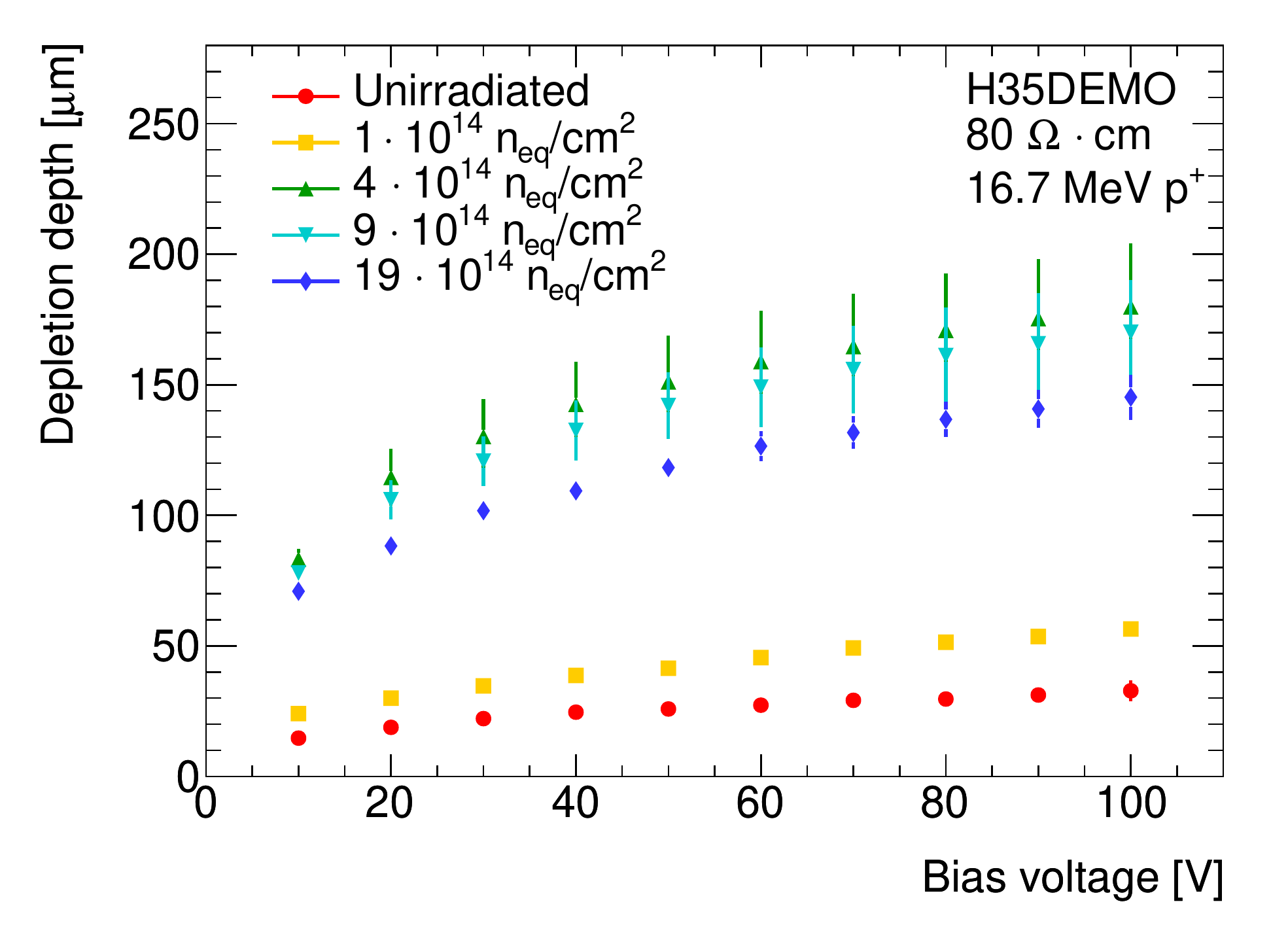}} \\
\subfloat[][\emph{}]
   {\includegraphics[width=.45\textwidth]{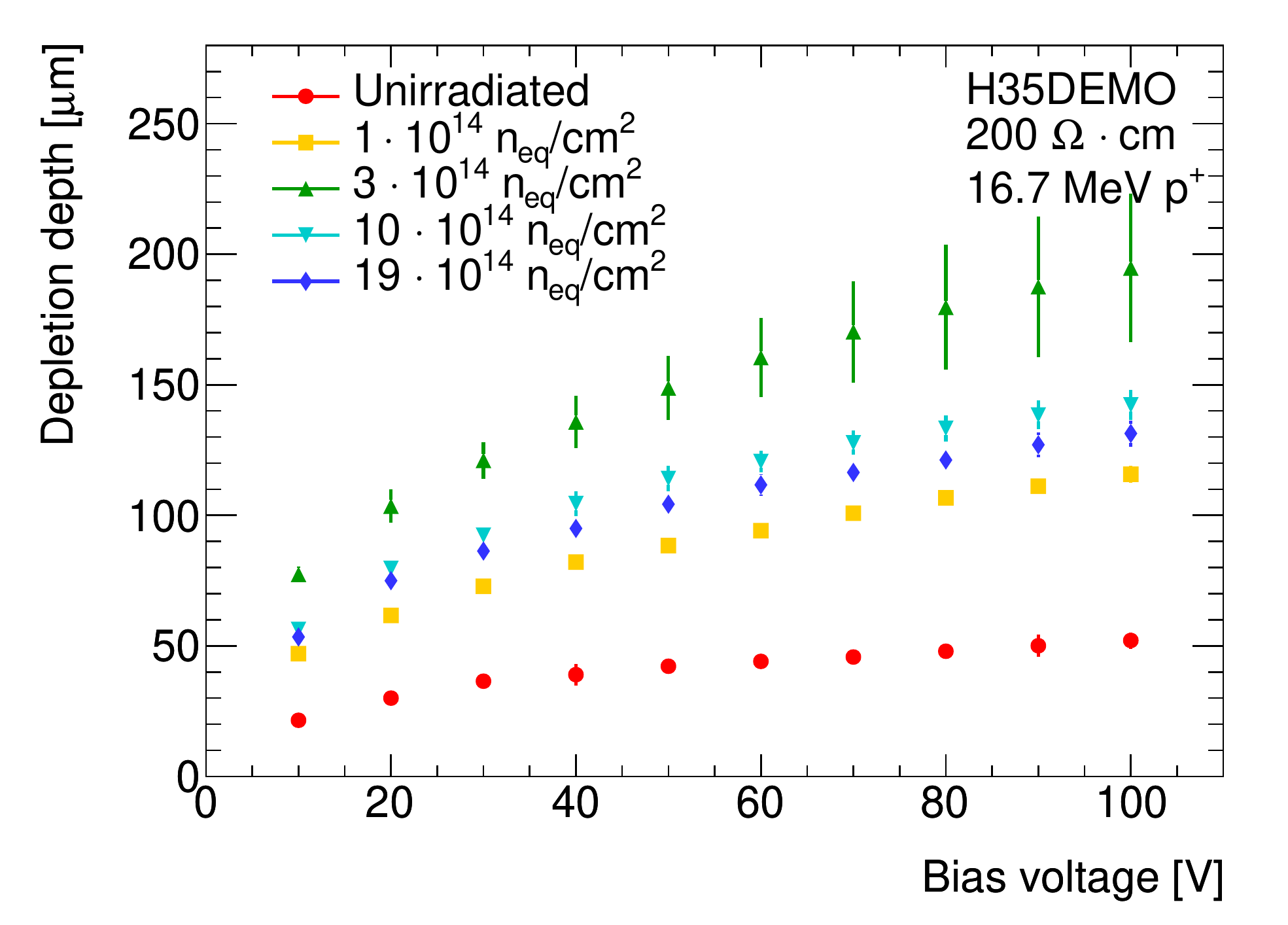}} \quad
\subfloat[][\emph{}]
   {\includegraphics[width=.45\textwidth]{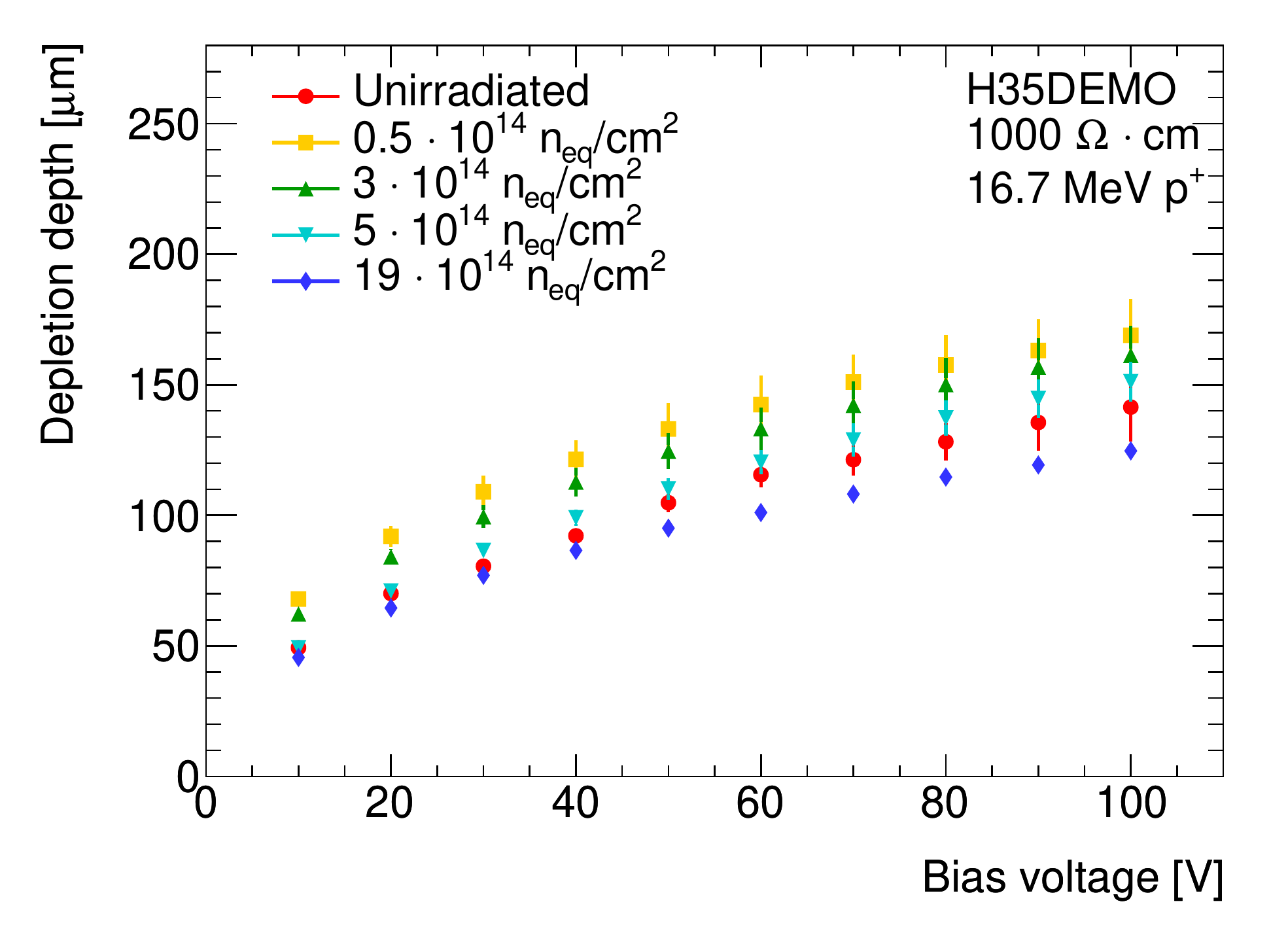}}
\caption{Evolution of the depletion depth as a function of the bias voltage, for the samples irradiated with protons of \SI{16.7}{\MeV}. The four plots show the behaviour of samples with different resistivities: 20, 80, 200 and 1000 \ohmcm{}}
\label{fig:dose_evolution_bern}
\end{figure}

\begin{figure}[t]
\centering
\subfloat[][\emph{}]
   {\includegraphics[width=.45\textwidth]{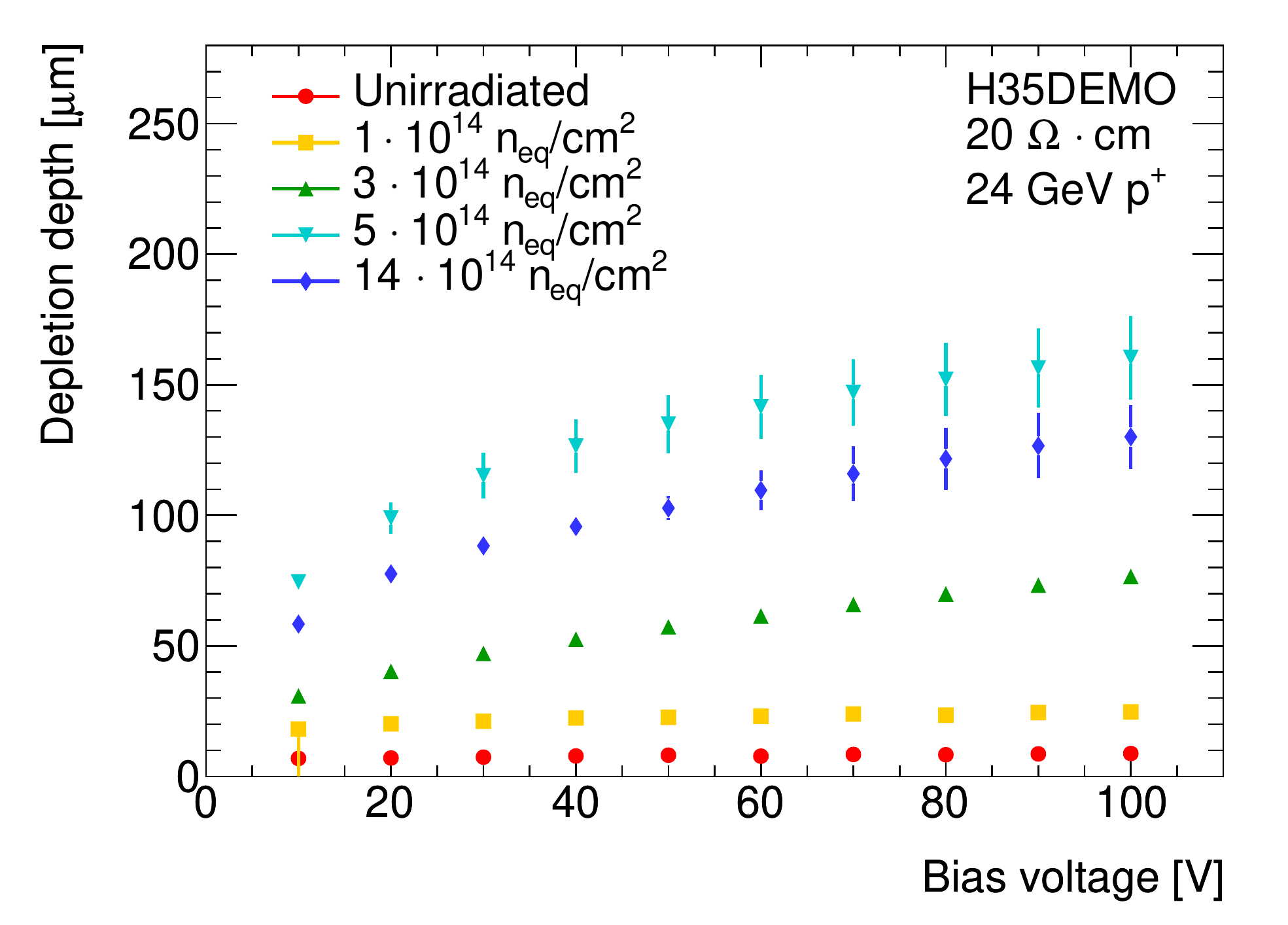}} \quad
\subfloat[][\emph{}]
   {\includegraphics[width=.45\textwidth]{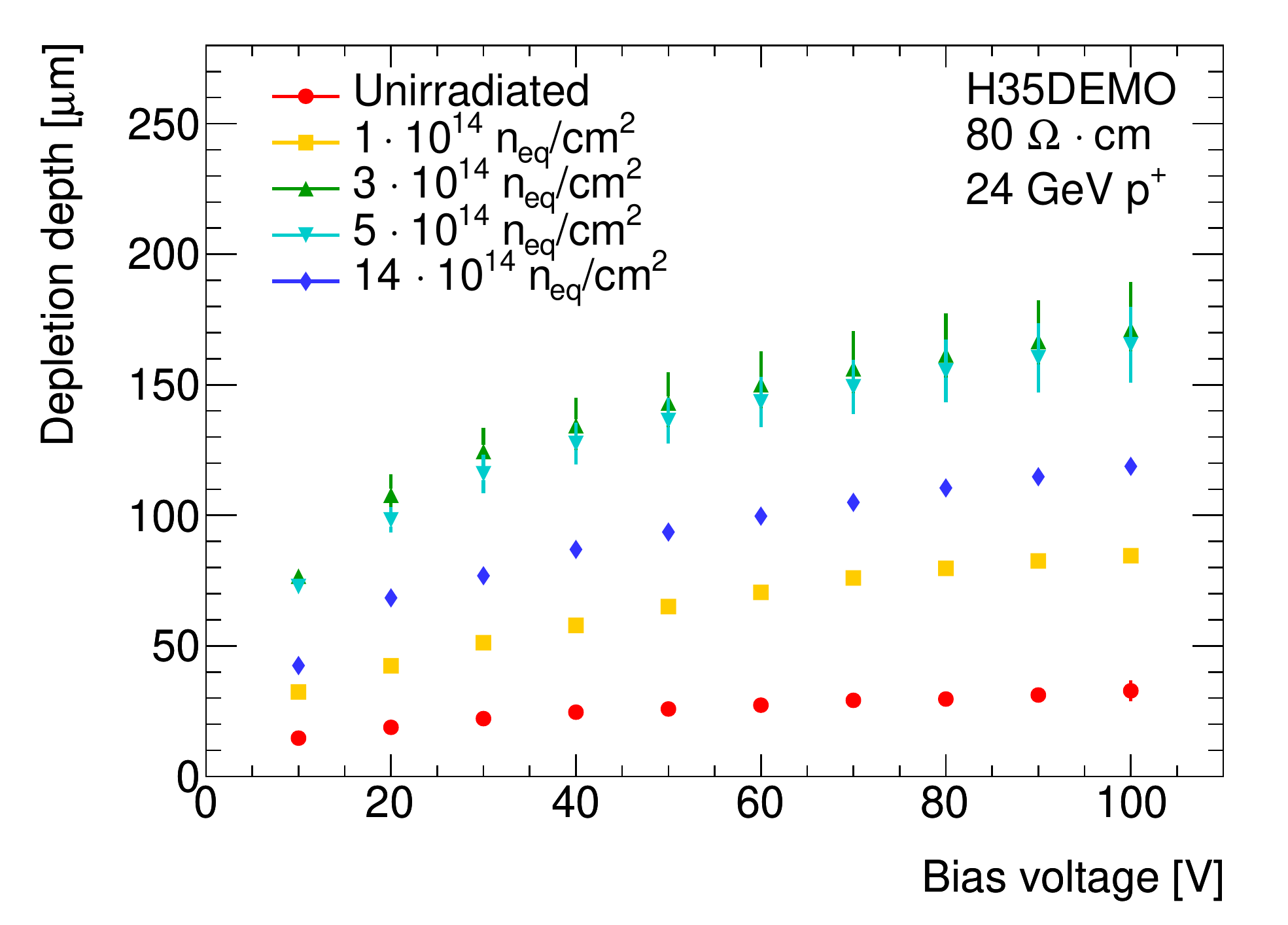}} \\
\subfloat[][\emph{}]
   {\includegraphics[width=.45\textwidth]{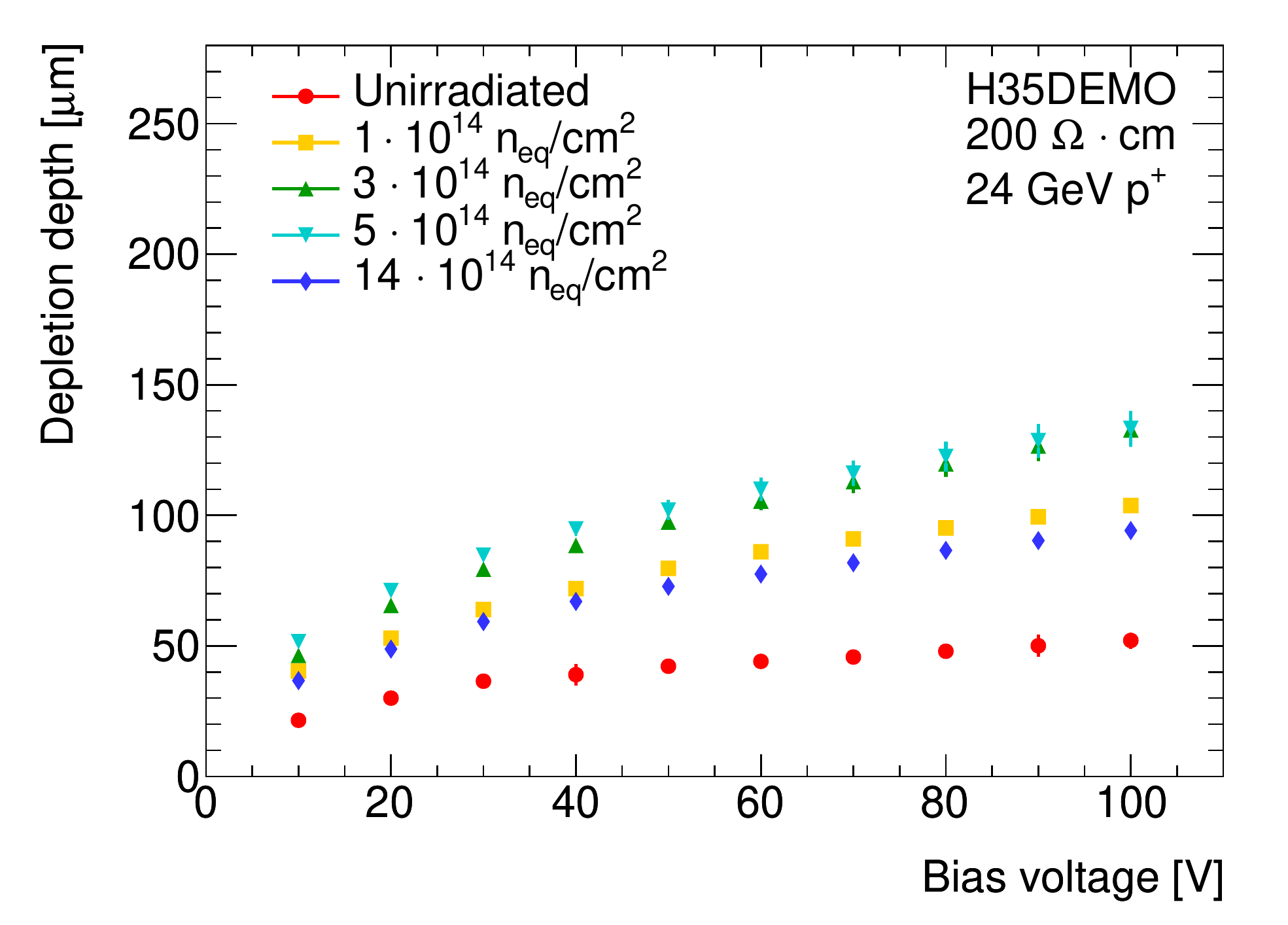}} \quad
\subfloat[][\emph{}]
   {\includegraphics[width=.45\textwidth]{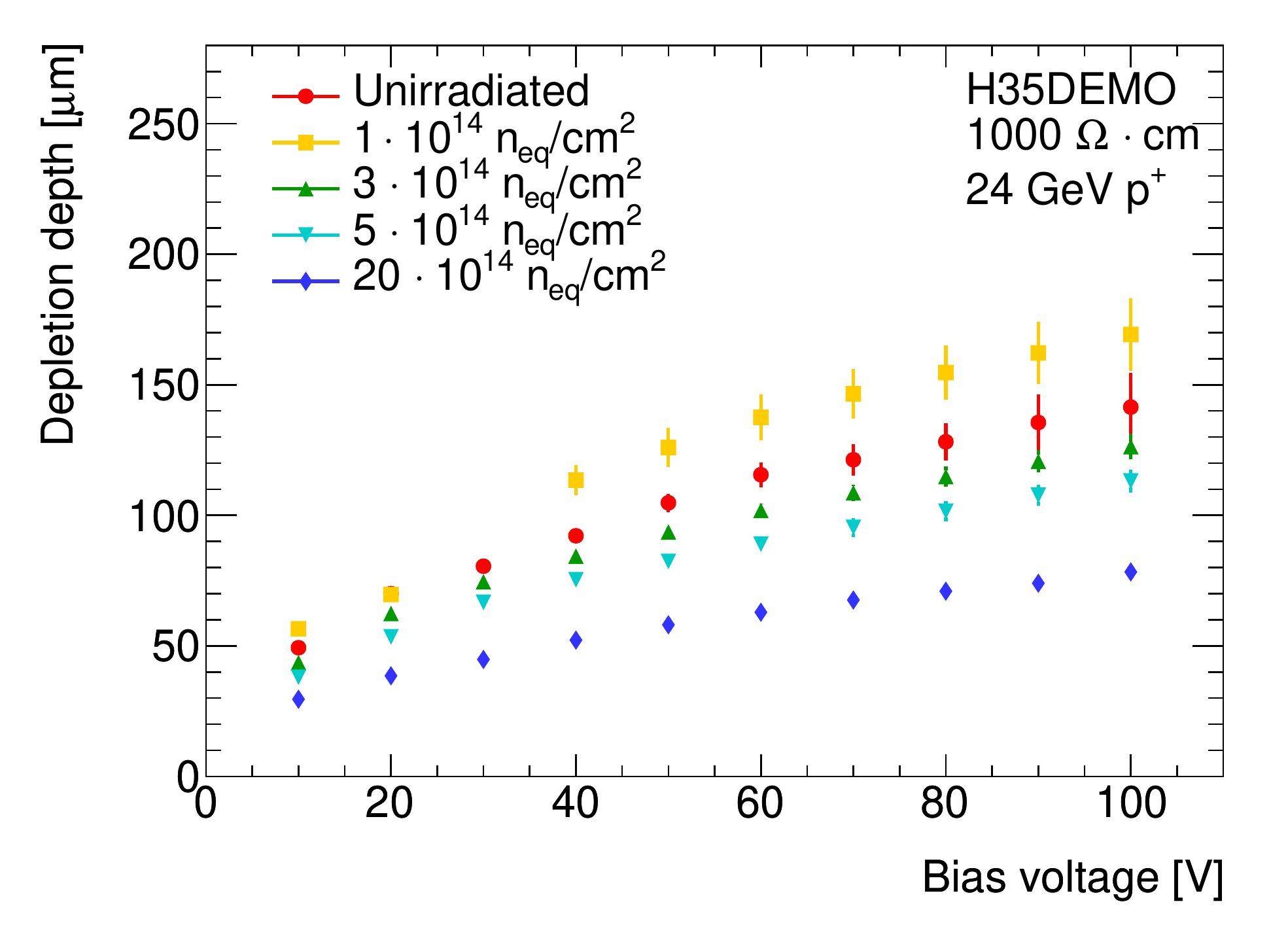}}
\caption{Evolution of the depletion depth as a function of the bias voltage, for the samples irradiated at \SI{24}{\GeV}. The four plots show the behaviour of samples with different resistivities: 20, 80, 200 and 1000 \ohmcm{}}
\label{fig:dose_evolution_PS}
\end{figure}

The samples were characterised after each irradiation step as described in Section~\ref{sec:non-irr}. The results of the measurement of the depletion region are shown in Figure~\ref{fig:dose_evolution_bern} for the samples irradiated with protons of \SI{16.7}{\mega\eV} and in Figure~\ref{fig:dose_evolution_PS} for the samples irradiated with protons of \SI{24}{\giga\eV}. 

For all samples, the depletion depth follows the same evolution with the increasing fluence: a strong growth is observed at low fluences, reaching values between \SI{140}{\um} and \SI{200}{\um}, depending on the resistivity, and then it shrinks again with the increase of the fluence. 
This effect has been observed before in sensors with p-doped substrates \cite{Mandic:2018afg, 1748-0221-11-02-P02016, 1748-0221-11-04-P04007} and can be explained using a combination of two different effects: proton irradiation causes the introduction of deep acceptors, which lead to a decreased depletion region, and, at the same time, shallow acceptors are removed. In this context the term ``shallow'' is used to indicate that the levels are close enough to the valence band to be fully ionised at room temperature.
At lower fluences the latter effect dominates, resulting in the observed initial increase of the active area while the former occurs at higher fluences. 

The maxima of the depletion region are reached at different fluence levels for the different resistivities, with higher resistivity samples having a faster evolution than the lower ones: the larger depletion depth is observed at a fluence of around \numprint{5e14}~\Neq{} for the 20~\ohmcm{} sample, while it is already reached before  \numprint{1e14}~\Neq{} in case of the 1000~\ohmcm{} resistivity.

Since the sensors are biased from the top, the charge is collected along non-straight field lines, and in particular when the larger depletion depths are obtained, the approximations considered in Equation~\ref{eq:rhofit} no longer hold and the fit fails to accurately describe the data. For this reason it was not possible to perform a quantitative study on the evolution of the resistivity.

By comparing the results obtained with protons of different energies, it is possible to observe that the evolution trend is similar, but the initial increase in depth due to the \SI{24}{\giga\eV} protons is less intense, and in general a smaller depletion region is measured with respect to the samples irradiated with \SI{16.7}{\mega\eV} protons.

\begin{figure}
\centering
\subfloat[][\emph{}]
   {\includegraphics[width=.45\textwidth]{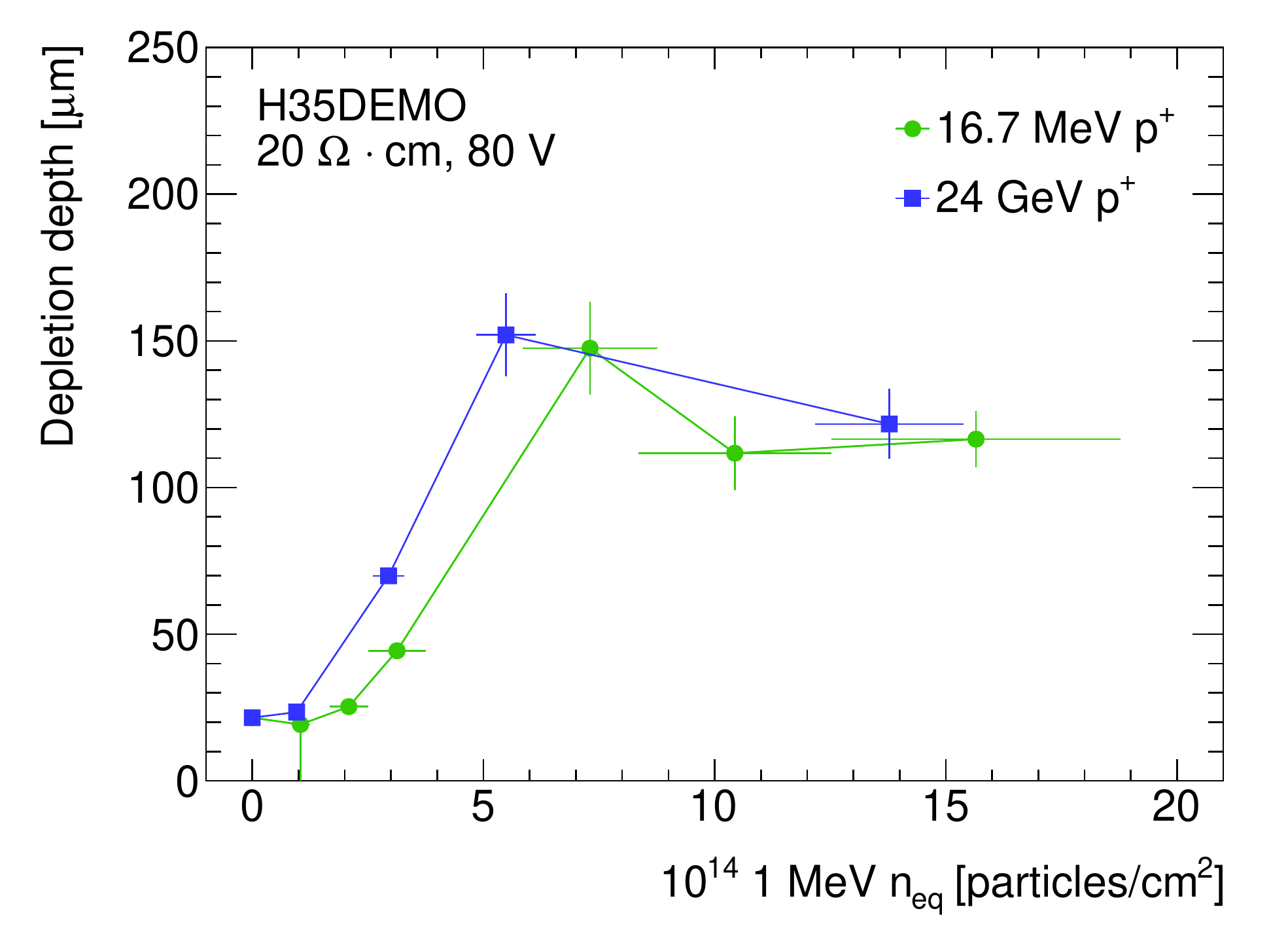}} \quad
\subfloat[][\emph{}]
   {\includegraphics[width=.45\textwidth]{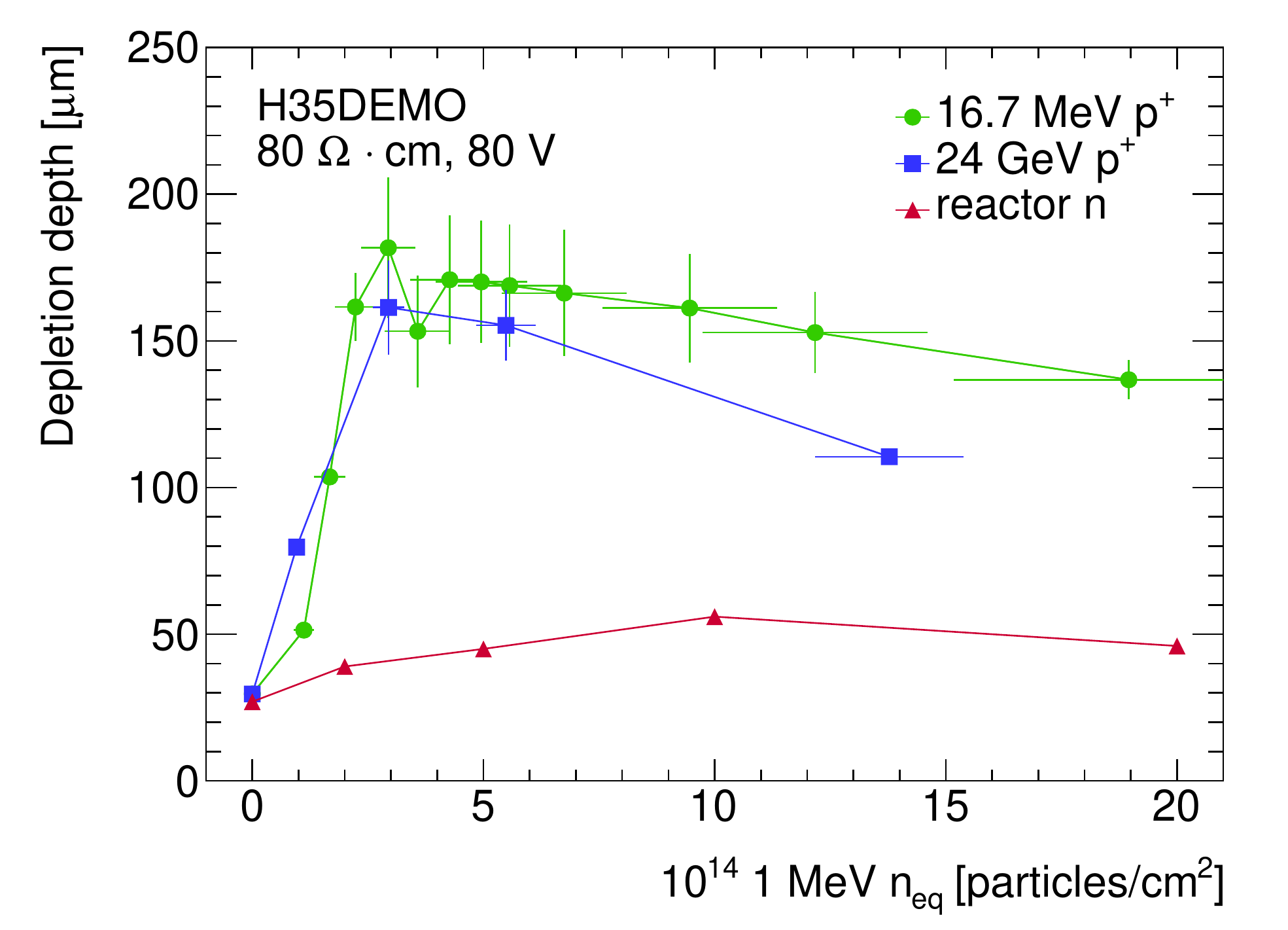}} \\
\subfloat[][\emph{}]
   {\includegraphics[width=.45\textwidth]{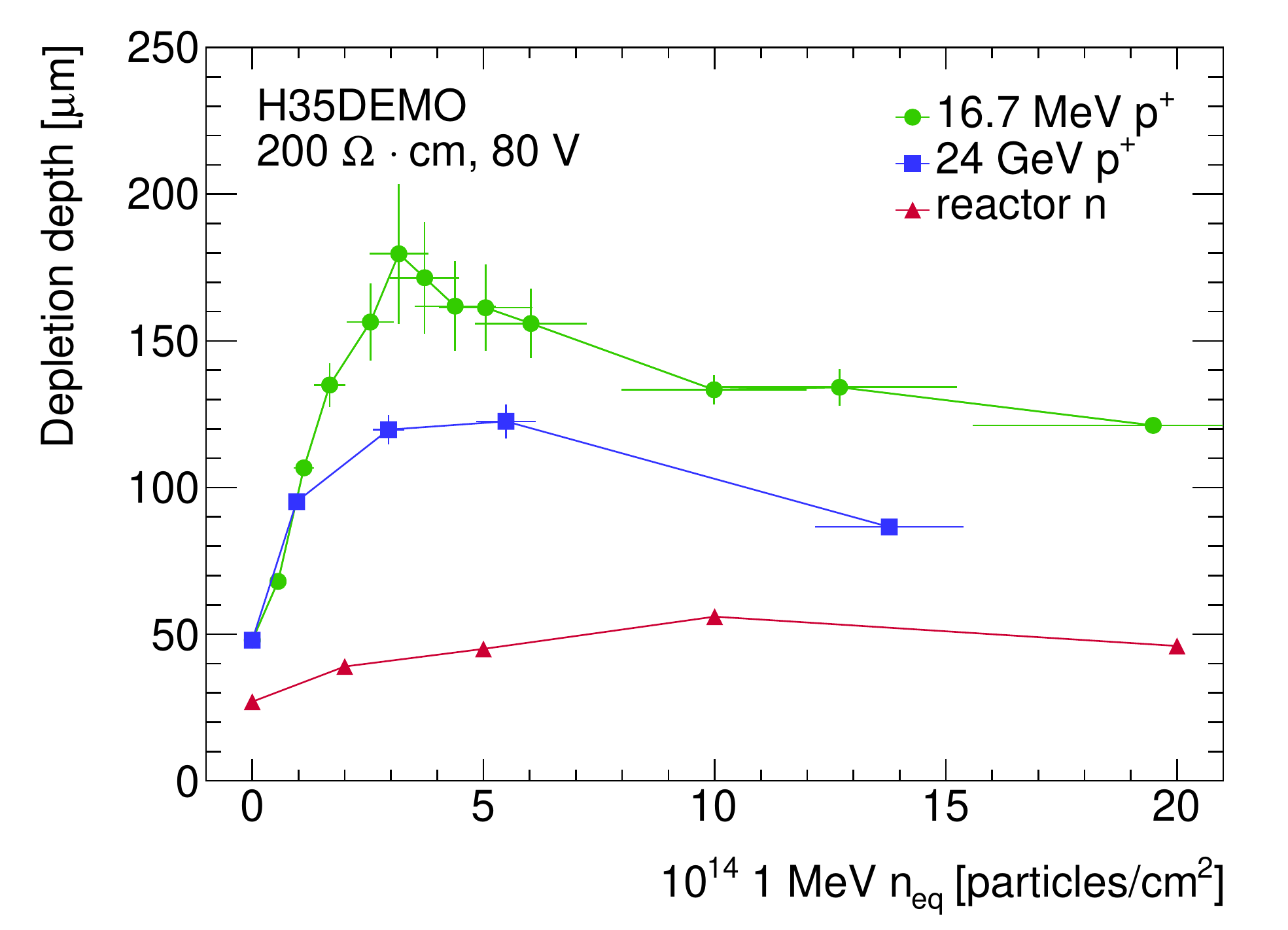}} \quad
\subfloat[][\emph{}]
   {\includegraphics[width=.45\textwidth]{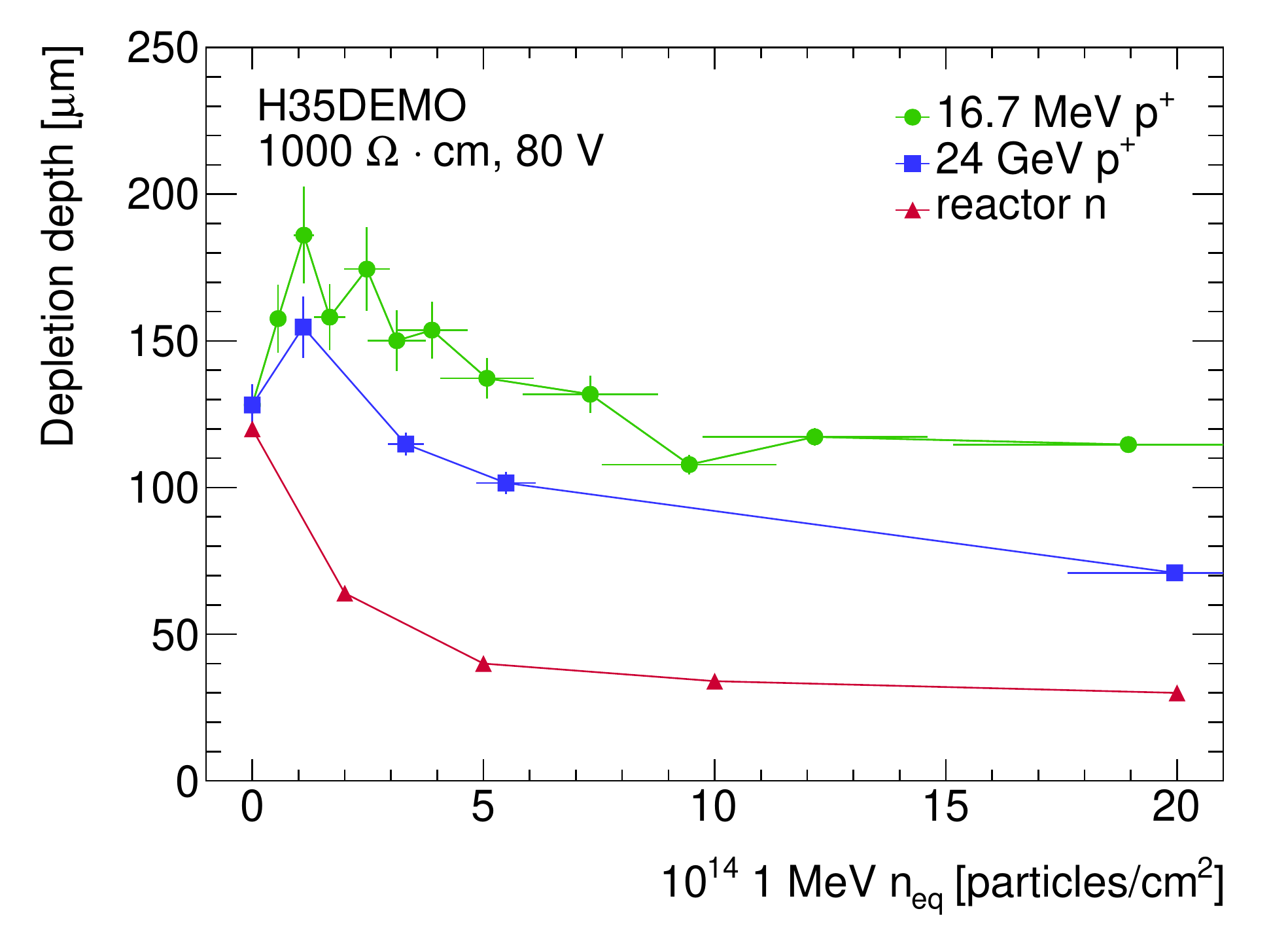}}
	\caption{Evolution of the depletion depth (at $-80$~V bias voltage) as a function of the fluence, comparing the damage arising from different particles: \SI{16.7}{\mega\eV} (circles) and \SI{24}{\giga\eV} (squares) protons and reactor neutrons (triangles). The neutron irradiation measurement are taken from \cite{ifae-tct}. The plots show the behaviour of samples with different resistivities: 20, 80, 200 and 1000~\ohmcm{}. }
	\label{fig:summary}
\end{figure}

The H35DEMO prototypes were also characterised with similar techniques after neutron irradiation, in the work presented in~\cite{ifae-tct}; this allows for an interesting comparison of the different effects of the non-ionising energy released by neutral and charged hadrons. Figure~\ref{fig:summary} shows the evolution of the depletion depth as a function of the fluence for a fixed reverse bias voltage ($80$~V), in the three irradiation campaigns: the \SI{16.7}{\mega\eV} and \SI{24}{\giga\eV} protons presented in this paper, and the nuclear reactor neutrons from~\cite{ifae-tct}. While the proton irradiated samples show a similar behaviour, the neutron samples show a lower increase of the depletion region for all the tested resistivities. This is due to the different nature of the interaction of the charged and neutral particles inside the silicon bulk.

	\section{Conclusions}
\label{sec:conclusions}

The depletion depth of the H35DEMO prototype has been measured before and after proton irradiation on samples of four resistivities.
All measured samples display a strong increase of the depletion depth after the initial irradiation steps, reaching a maximum between \numprint{1e14} and \numprint{3e14}~\Neq{}, when irradiated with 16.7~MeV protons, and between \numprint{1e14} and \mbox{\numprint{6e14}~\Neq{}} when irradiated with 24~GeV protons. The increase of the depletion depth at low fluence can be attributed to the removal of shallow acceptors. The depletion depth is then measured to decrease slowly with the increasing fluence.

This work shows a striking different effect in the charge collection properties after the irradiation of a p-type silicon bulk with neutrons and protons.

Comparison of these results with the measurements obtained after neutron irradiation \cite{ifae-tct} shows a strong difference in the effect of the two particles. While the depletion region exceeds 100~\si{\micro m} after proton irradiation for all the resistivities, it does not exceed 60~\si{\micro m} in the case of neutron irradiation.
Furthermore, the growth of the depletion region after neutron irradiation continues up to higher fluences compared to proton irradiation, for the 80 and 200~\ohmcm{} samples.
The differences between protons with different energies are not as pronounced and the trend is similar. Still, the effect of high energy protons is smaller than the low energy ones.


\acknowledgments

This project was funded by the European Union's Horizon 2020 research and innovation program under the Marie Skłodowska-Curie grant agreement No 675587, and by the SNSF grants 20FL20\_173601, 200021\_169015 and 200020\_169000. The authors gratefully acknowledge the support by the
mechanical and electronic workshop of the DPNC (Universiy of Geneva) and LHEP (University of Bern).



\bibliographystyle{BibStyle}
\bibliography{bibliography.bib}

\end{document}